\RequirePackage{rotating}
\documentclass[a4paper,fleqn,usenatbib]{mnras}
\usepackage{newtxtext,newtxmath}
\usepackage[T1]{fontenc}
\usepackage{ae,aecompl}
\usepackage{ulem}  

\usepackage{eso-pic}

\AddToShipoutPictureBG*{%
  \AtPageUpperLeft{%
    \hspace{0.75\paperwidth}%
    \raisebox{-4.3\baselineskip}{%
      \makebox[0pt][l]{\textnormal{DES-2022-0727}}
}}}

\AddToShipoutPictureBG*{%
  \AtPageUpperLeft{%
    \hspace{0.75\paperwidth}%
    \raisebox{-5.2\baselineskip}{%
      \makebox[0pt][l]{\textnormal{FERMILAB-PUB-23-508-PPD}}
}}}%

\usepackage{graphicx}	
\usepackage{amsmath}	
\usepackage{lscape} 
\usepackage{rotating}
\usepackage{lineno}

\newcommand*{\logTen}{\ensuremath{\log_{10}}}

\newcommand{\fcont}{{$f_\mathrm{cont}$}}

\newcommand{\Planck}{\textit{Planck} }

\title[The SPT-SZ MCMF cluster catalog]{SPT-SZ MCMF:  An extension of the SPT-SZ catalog over the DES region}

\author[Matthias Klein et al.]{
\parbox{\textwidth}{
\Large
M.~Klein,$^{1}$\thanks{E-mail:matthias.klein@physik.lmu.de}
J.~J.~Mohr,$^{1,2}$
S.~Bocquet,$^{1}$
M.~Aguena,$^{3}$
S.~W.~Allen,$^{4,5,6}$
O.~Alves,$^{7}$
B.~Ansarinejad,$^{8}$
M.~L.~N.~Ashby,$^{9}$
D.~Bacon,$^{10}$
M.~Bayliss,$^{11,12}$
% ...
}
\vspace{0.4cm}
\\
\underline{\normalsize \em Affiliations at the end of the paper.}
}

\date{Accepted XXX. Received YYY; in original form ZZZ}
\pubyear{2023}
\begin{document}
\label{firstpage}
\pagerange{\pageref{firstpage}--\pageref{lastpage}}
\maketitle


\begin{abstract}
We present an extension to a Sunyaev-Zel'dovich Effect (SZE) selected cluster catalog based on observations
from the South Pole Telescope (SPT);  this catalog extends to lower signal-to-noise than the previous SPT-SZ catalog and therefore includes lower mass clusters. Optically derived redshifts, centers, richnesses and morphological parameters together with catalog contamination and completeness statistics are extracted using the multi-component matched filter algorithm (MCMF) applied to the S/N$>$4 SPT-SZ candidate list and the Dark Energy Survey (DES) photometric galaxy catalog. 
The main catalog contains 811 sources above S/N=4, has 91\% purity and is 95\% complete with respect to the original SZE selection.
It contains 50\% more total clusters and twice as many clusters above $z=0.8$ in comparison to the original SPT-SZ sample. The MCMF algorithm allows us to define subsamples of the desired purity with traceable impact on catalog completeness. As an example, we provide two subsamples with S/N$>$4.25 and S/N$>$4.5 for which the sample contamination and cleaning-induced incompleteness are both as low as the expected Poisson noise for samples of their size. The subsample with S/N$>$4.5 has 98\% purity and 96\% completeness, and will be included in a combined SPT cluster and DES weak-lensing cosmological analysis. 
We measure the number of false detections in the SPT-SZ candidate list as function of S/N, finding that it follows that expected from assuming Gaussian noise, but with a lower amplitude compared to previous estimates from simulations.

\end{abstract}



\section{Introduction}


Within the past ten to twenty years, cluster catalogs based on the detection of ICM either via its X-ray emission or via the Sunyaev-Zel'dovich Effect (SZE) signature have grown from just tens of systems to thousands of systems \citep{Lahav89,NORAS,Vanderlinde10,Bleem15,Bleem20,CODEX19,ACTDR5,Klein23} and will soon reach tens or even hundreds of thousands \citep{Merloni12,Raghunathan22}.
These ICM selected cluster catalogs require optical follow-up to assign cluster redshifts and typically to confirm that the ICM-selected cluster candidate is physically associated with a cluster of galaxies.

With the availability of well-calibrated, large solid-angle optical surveys \citep[e.g., SDSS, KIDS, DES, HSC-SSP, Legacy Surveys;][]{SDSS,deJong13,Flaugher15,HSC-SSP,Legacysurveys19} and mid-infrared surveys like that from the Wide-Field Infrared Survey Explorer \citep[WISE,][]{WISE},
 the confirmation and redshift assignment can be done systematically over large portions of the sky. In the past, the final cluster catalogs--- especially those employed for cosmology--- were often defined such that the confirmation and redshift assignment would have a negligible impact on the completeness of the original ICM candidate list \citep[e.g.,][]{Mantz10,Benson13,Reichardt13,Bocquet15,dehaan16}. This approach is now coming to its limit, because the larger cluster samples needed for improved cosmological constraints require improved control over systematics-- including even the impact of follow-up confirmation and redshift assignment on sample completeness. Moreover, using only information from the ICM-selected catalog to produce a clean sample will lead to significantly smaller samples than would be possible using additional information from the optical follow-up.

Examples of combining X-ray-selected cluster candidate catalogs and systematic optical follow-up include the confirmation of ROSAT selected clusters via DES \citep{Klein19}, SDSS \citep{CODEX19} and the Legacy Survey DR10 \citep{Klein23}.  These efforts yielded thousands of new galaxy clusters extending to higher redshifts and with an angular density many times higher than previously selected ROSAT cluster samples that relied on individual cluster imaging and spectroscopy. 
The eFEDS X-ray survey \citep{Brunner21} carried out by eROSITA on the satellite Spektrum-Röntgen-Gamma \citep{Predehl21} has been analyzed with MCMF using HSC-SSP and Legacy Survey DR9 data yielding a 94\% pure sample with 477 confirmed clusters over 140 deg$^2$. A subset with 450 clusters was recently used for the first eROSITA-based cluster cosmology \citep{Chiu23}. The usage of MCMF based cleaning in this study allowed us to increase the sample useful for the cosmological study by more than a factor two compared to solely relying on X-ray data.

Similar systematic optical follow-up of SZE-selected cluster candidates has been carried out. The analysis of a set of SPTpol-ECS candidates was pursued with the redMaPPer algorithm \citep{rykoff14} in targeted mode using DES data, supplemented with WISE and Panstarrs survey and pointed Spitzer IR and PISCO observations \citep{Bleem20}.  The ACT cluster candidate list has also been systematically followed up using DES and other data \citep{ACTDR5}. Recently, a new low S/N SZE-selected candidate list from the \Planck mission dataset has been followed up using the MCMF algorithm with DES data, resulting in the discovery of the highest redshift \Planck selected systems to date as well as a tripling of the number of confirmed \Planck clusters in the DES survey footprint \citep{Hernandez22}.

In the analysis presented here, we carry out a similar study of confirming an SPT-SZ candidate list that extends to lower S/N than has been previously attempted in \citet{Bleem15}.  We apply the MCMF algorithm to the SPT-SZ candidate list down to S/N=4 using the DES and WISE datasets, cross-checking previously confirmed SPT-SZ clusters but also identifying many lower-mass, previously undiscovered galaxy clusters.

This paper is structured as follows. In Section~\ref{sec:data} we describe the dataset used in this work, and in 
Section~\ref{sec:method} we outline the cluster confirmation method.  The SPT-SZ MCMF cluster catalog is presented in Section~\ref{sec:SPT-SZ MCMF} and validated in Section~\ref{sec:validation}. The conclusions are summarized in Section~\ref{sec:conclusions}. 
Throughout this paper we adopt a flat $\Lambda$CDM cosmology with $\Omega_M=0.3$ and $H_0=70$\,km\,s$^{-1}$\,Mpc$^{-1}$.


\section{Data}
\label{sec:data}
In this paper we make use of the photometric catalog from DES observations obtained within the first three years of the survey (Y3) and the SPT-SZ cluster candidate list down to S/N=4. For the high-z confirmation of cluster candidates, we further make use
of mid-IR data from the WISE satellite \citep{WISE,Mainzer11} in the form of a matched catalog between DES and the UnWISE catalog \citep{unWISE3cat}.
The following subsections provide an overview of the datasets used.

\subsection{The DES Y3A2 GOLD catalog}

For the optical confirmation out to $z\approx1.3$ we make use of the DES Y3A2 GOLD catalog, which is based on $g$, $r$, $i$ and $z$ band DECam \citep{Flaugher15} imaging data from DES between August 2013 and February 2016. Details on the data reduction and data quality are given elsewhere \citep{DESDR1,Morganson18}.

The DES Y3A2 GOLD catalog is a value-added version of the photometric catalog released in the public data release 1 \citep[DR1;][]{DESDR1}.
The catalog covers approximately 5000\,deg$^2$ in area with typically 3-5 exposures per band and reaches  95\% completeness limits of  23.72, 23.34, 22.78 and 22.25~mag in the $g$, $r$, $i$ and $z$ bands, respectively.
The catalog includes additional calibration steps, flags and types of photometry. In our work we make use of the multi-epoch, multi-band,  multi-object fitting photometry "MOF", which  is based on the ngmix code \citep{Sheldon14} and fits a galaxy model to each 
single epoch exposure and band at the same time, considering the different PSF shapes and sizes. Furthermore, it simultaneously fits neighbouring sources for improved deblending.  In addition to MOF we make also use of single-object fitting (SOF) photometry, which is derived in a similar way but masking neighbouring sources rather than fitting them. As SOF turned out to be more robust, while MOF provides better photometry in crowded regions, we make use of SOF photometry in cases where MOF has failed.

Out to $i=22.2$~mag we use the star-galaxy separator available in GOLD, which is an expanded version of that available in DES Y1A1 \citep{DESY1Gold} and includes MOF/SOF-based extent information.  For fainter magnitudes we do not apply a star-galaxy separation to maximise sensitivity to small, high-redshift cluster galaxies.
The resulting impact on cluster richness from residual contamination by stars in the galaxy sample is minimized by using a local background measurement, which works well in the limit that the residual stellar density near the cluster position is nearly constant.

In addition, we make use of mask flags to exclude regions around bright stars and the "top of the galaxy" calibration including SED-based de-reddening of sources due to interstellar dust provided in the DES Y3A2 GOLD catalog.

\subsection{The SPT-SZ cluster candidates with S/N>4}\label{sec:spt}
The SPT-SZ survey is based on observations with the SPT-SZ camera on the 10m South Pole Telescope \citep[SPT;][]{Carlstrom11}, which has a 1~degree diameter field-of-view and a resolution of about $\sim$1~arcmin.
The survey was conducted from 2007 to 2011, covering 2,500~deg$^2$ between 20h<RA<7h and -65\textdegree <Dec<-40\textdegree and in three frequencies 95, 150 and 220~GHz.
The source detection via the thermal SZE 
is performed on the 95 and 150~GHz maps using a matched-filter approach \citep{Bleem15}.
The SPT-SZ cluster candidate list contains 1,518 sources with S/N$>$4, of which 1,395 (92\%) fall within unmasked areas of DES that are suitable for optical follow-up.

\subsection{WISE}\label{sec:unwise}
The WISE satellite is a mid-infrared telescope with a main mirror of 40 cm observing in four bands at 3.4 \textmu m, 4.6 \textmu m, 12 \textmu m and 22 \textmu m ($w1$, $w2$, $w3$, $w4$).
The observing campaign can be divided in three phases the main phase, with sufficient cooling propellant to observe the full-sky 1.5 times in all four bands. A second phase called NEOWISE was performed
immediately after the main campaign and without cooling completing the second full-sky observations in the $w1$ and $w2$. A third phase of WISE observations (NEOWISE-R) started in September 2013 when WISE was recommissioned after more than two years of hibernation. Since then WISE completes a full-sky survey every $\sim 6$ months.

In this work we use the unWISE catalog \citep{unWISE19} that makes use of all WISE data until the end of the first year of the NEOWISE-R phase. It is based on the unblurred coadds of WISE imaging data \citep[unWISE][]{unWISE14} and includes improved source detection and deblending modeling for crowded regions. The catalog yields a gain of 0.7 mag in depth and contains twice the number of galaxies with respect to the AllWISE catalog \citep{Cutri13} that is based on solely the main and he NEOWISE phase of WISE observations.


\section{Cluster Confirmation Method}
\label{sec:method}

For cluster confirmation and redshift determination of the majority (>90\%) of SPT-SZ cluster candidates we use the multi-component matched filter cluster confirmation tool \citep[MCMF; see details in][]{Klein18,Klein19} with DES photometric data. In Section~\ref{sec:MCMF} we summarise the method and describe some recent modifications.

From the previous SPT-SZ sample \citep{Bleem15} we expect a significant fraction ($\sim 8$\%) to be at $z>1$, where the DES imaging data need to be complemented with NIR or IR imaging.
For that reason we develop a high-z cluster confirmation tool, following the MCMF concept but using a combination of DES and WISE (mid-IR) photometry data.  This is described in Section~\ref{sec:HIGHZ}.  Finally, we review the optical morphological measures that we extract for the SPT-SZ MCMF clusters in Section~\ref{sec:dynstate}.

\subsection{MCMF}
\label{sec:MCMF}
The MCMF algorithm has been designed for the confirmation and characterization of ICM-selected cluster candidates identified in large X-ray or SZE surveys. MCMF has been successfully applied to ROSAT X-ray sources over the DES footprint \citep[MARD-Y3;][]{Klein19} and more recently in combination with the Legacy Survey DR10 dataset \citep{Legacysurveys19}, it has been used to create the all-sky optically-confirmed X-ray cluster catalog \citep[RASS-MCMF;][]{Klein23}.  In addition, it was used for the optical follow-up of the first eROSITA-based galaxy cluster catalog over the early mission test field eFEDS \citep{Klein22}.  Beyond this, it has also been applied to new S/N$>$3 Planck SZE-selected catalogs over the DES region \citep[MADPSZ][]{Hernandez22}.  In working with these different datasets, improvements and extensions to the original method have been made.  In these applications, the new MCMF based catalogs significantly enhanced the number of clusters that had been previously extracted from the same X-ray or SZE datasets and followed up with cluster-by-cluster imaging and spectroscopy.  In addition to enlarging the samples, the MCMF method allows one to limit the contamination of the new samples.

 The MCMF algorithm includes a red sequence technique \citep{gladders00,rykoff14} with redshift and magnitude dependent color filters in the $g-r$, $r-i$ and $i-z$ colors, a radial weighting (projected NFW profile centered at ICM selected candidate location) and a characteristic magnitude range to estimate redshifts and richnesses for candidates. From the cluster candidate list it makes use of the source position and an ICM-based mass proxy. The mass proxy is used to estimate the radius $R_{500}$ within which galaxies are counted to estimate cluster richness. In this work we make use of the SPT-SZ candidate S/N together with a calibration of the S/N-to-mass relation \citep{Bocquet19} to extract a cluster mass estimate for a range of hypothetical redshifts.

For each cluster candidate, the color and radially weighted, background-subtracted richness $\lambda(z)$ within $R_{500}$ is calculated as a function of (a priori unknown) candidate redshift.
The peaks in $\lambda(z)$ are then identified and modeled with so-called ``peak profiles'' (see below). 
If present, multiple richness peaks ($\le3$) along the line of sight toward each candidate are recorded. Examples of peak profiles and their best fit to $\lambda(z)$ profiles of clusters are presented in previous MCMF analyses \citep[e.g. Figs. 4 \& A2,][]{Klein19}.  These peaks with associated richnesses and redshifts are then collected and processed further as described in the following subsections.

Note that the peak profile models are built using renormalised stacks of individual $\lambda(z)$ profiles from clusters with spectroscopic redshift measurements (spec-z's). 
The clusters with spectroscopic redshift do not need to be
part of the sample that is being studied.
Important here is that the redshift dependency of the SZE observable-derived estimate of $R_{500}$ be the same for the spec-z clusters and the candidates to be analysed. To ensure this, we assign a value $\xi$ of the SZE observable S/N to the spec-z clusters that is consistent with their masses and redshifts that then can be used as input to the MCMF pipeline. 

To confirm clusters we characterize the likelihood that a given optical counterpart is a chance superposition rather than a physical counterpart to the ICM-selected cluster candidate.  Doing so requires knowing the typical richness distribution of contaminants as a function of redshift within the survey region.
Thus the same exercise employed on the candidates is then repeated using random positions within the SPT-SZ footprint, using the same distribution of $\xi$ and excluding regions containing SPT-SZ detections. These random positions provide the richness distribution of non-SZE detected structures (noise, projections, undetected clusters).  The richness distributions from the random lines-of-sight and true clusters are redshift dependent, because they are impacted by the selection function, the evolution of the halo mass function and the noise in the richness estimate.

To be able to control the contamination of the final cluster sample, we calculate a quantity \fcont.  High values of \fcont\ indicate a higher probability that the candidate in question is a chance superposition rather than a real cluster.  \fcont is calculated using the mean richness distributions along the random lines-of-sight $f_\mathrm{rand}(\lambda,z)$ and the richness distributions $f_\mathrm{obs}(\lambda,z)$ towards the candidates.  
That is, for each candidate $i$ we calculate the number of random lines-of-sight within a redshift bin with richness $\lambda\ge\lambda_i$ and divide by the number of SZE candidates within the same redshift bin with  $\lambda\ge\lambda_i$. This ratio is then re-scaled according to the total number of SPT candidates and random lines-of-sight.
\begin{equation}\label{eq:fcont}
   f_{\mathrm{cont}}(\lambda_i,z_i)=\frac{\int_{\lambda_i}^{\infty} f_\mathrm{rand}(\lambda,z_i) d\lambda}{\int_{\lambda_i}^{\infty}
f_\mathrm{obs}(\lambda,z_i) d\lambda},
\end{equation}
This \fcont\ parameter is calculated for each richness peak associated with a candidate.  The peak showing the lowest value of \fcont\ is assigned as the best optical counterpart for the SPT-SZ candidate, because it is the most likely to be a real cluster. 

The cluster sample itself can then
be defined as those candidates showing an \fcont\ below a certain threshold value $f_\mathrm{cont}^\mathrm{max}$. The threshold value corresponds to the fraction of the contamination in the initial candidate list that makes it into the final cluster catalog.
The contamination of the resulting final cluster catalog would then be $f_\mathrm{SZE-cont} \times f_\mathrm{cont}^\mathrm{max}$, where the confirmed catalog contains all candidates with $f_{\mathrm{cont}}(\lambda_i,z_i)\le f_\mathrm{cont}^\mathrm{max}$ and the initial contamination of the candidate catalog is $f_\mathrm{SZE-cont}$.
 As an example, if the input catalog is known to be 50\% pure and an \fcont\ threshold value $f_\mathrm{cont}^\mathrm{max}=0.2$ is employed, then the contamination fraction of the confirmed cluster catalog would be $0.5\times 0.2=0.1$ or 10\%.

The version of MCMF applied here is-- aside from the different mass proxy-- largely the same as the version applied previously to 
two previous X-ray samples \citep{Klein19,Klein22}. Some minor improvement we made on the estimate of the redshift uncertainty.
Based on the analysis on mock data, that include effects such as scatter in photometric calibration, intrinsic and measurement scatter of cluster member galaxy colors and structures along the line of sight, we find that cluster photo-z scatter can be reasonably well described as $\sigma_z = f(z)/\sqrt(\lambda)$. Here $f(z)$ is a scale factor as function of cluster redshift that is calibrated empirically with spectroscopic redshifts. The photometric redshift uncertainties that we list are therefore redshift and richness dependent.
A second improvement specific to this work is a second iteration on the estimate of the richness distributions along random lines-of-sight. The richness distributions along random lines-of-sight $f_\mathrm{rand}$ is supposed to resemble the expected richness distribution of contaminants as function of redshift. Given the correlation between $\xi$, cluster mass and likelihood of a source being a real cluster, the initial choice of using the same $\xi$ distribution as the full candidate list causes the estimate of \fcont\ to be mildly biased high. To avoid this bias we use the $\xi$ distribution of rejected systems (\fcont$>0.3$) as proxy of the distribution of contaminants and to select a subsample of randoms that follows this distribution and remeasure \fcont\ for all candidates.

 \begin{figure}
\includegraphics[keepaspectratio=true,width=0.97\columnwidth]{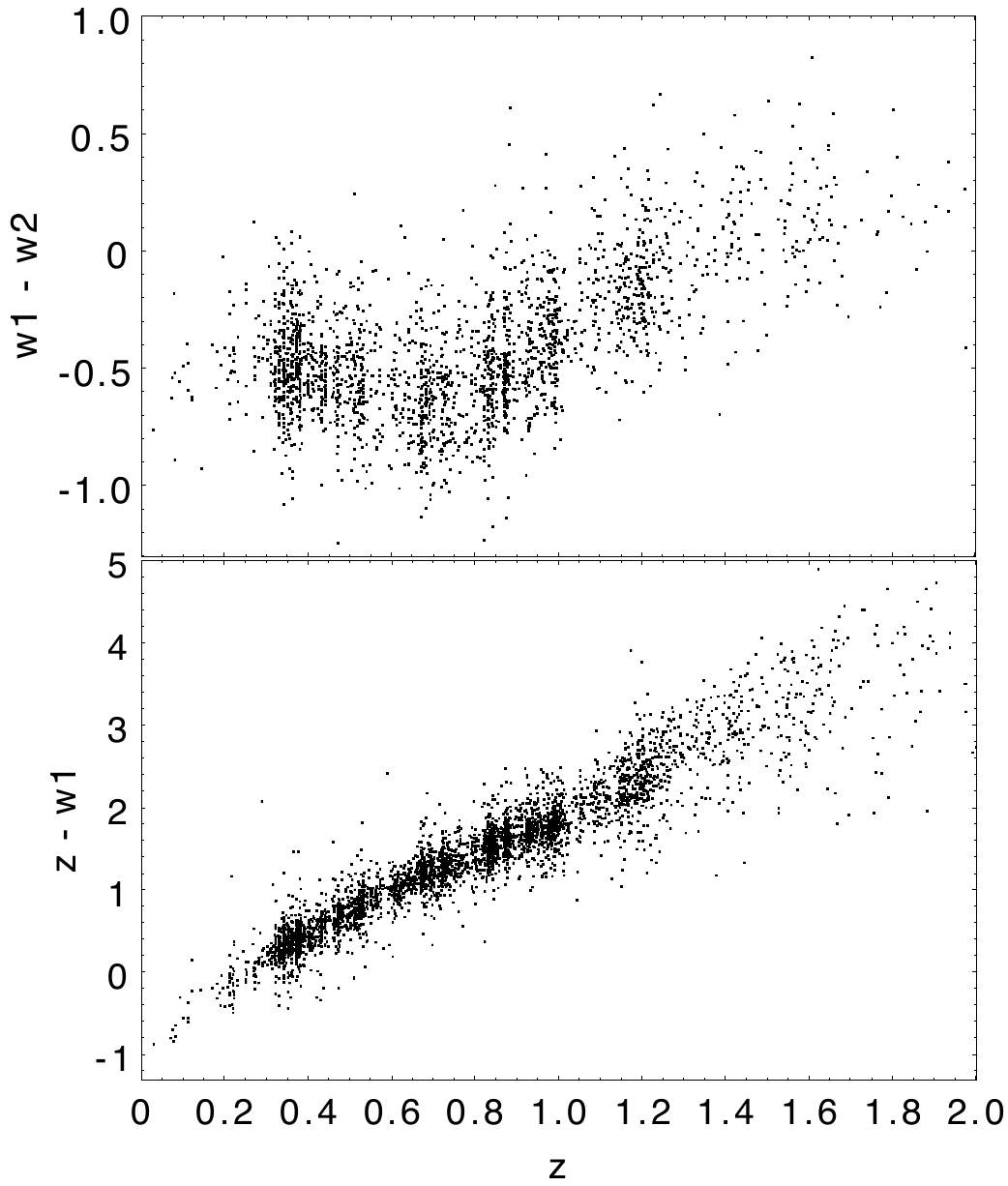}
\vskip-0.10in
\caption{Passive galaxy colors versus redshift in the COSMOS field, including the $w1$-$w2$ color from the unWISE catalog (top), and the the DES z minus WISE $w1$ color (bottom). The observed galaxy colors suggest that cluster redshift constraints can be obtained out to $z\approx1.5$ when adding WISE data to the DES data set.}
\label{fig:colordep}
\end{figure}

\subsection{High-redshift extension using WISE}
\label{sec:HIGHZ}
Besides the fact that passive galaxies become fainter with increasing redshift, the rest-frame wavelength range covered by the DES bands no longer brackets the 4000\,\AA \ break at redshifts $z\gtrsim1$. Therefore, photometric redshifts become increasingly uncertain at these redshifts.
For high-redshift cluster confirmation and photometric redshifts, it is therefore advantageous to move to redder bands such as the mid-IR regime covered by the Spitzer or WISE satellites. Data from both satellites were previously used for high-redshift cluster searches \citep{Muzzin09,Madcows} as well as for cluster confirmation \citep{Bleem15,Bleem20}.

In our current analysis, we use the unWISE catalog \citep{unWISE3cat} that additionally includes more recent $w1$ and $w2$ band WISE imaging data from the NEOWISE-R phase to create deep catalogs without PSF scale smoothing of the data and includes an improved modeling of crowded regions \citep{unWISE19}. WISE data exist over the full sky, and therefore we match 
the unWISE and DES catalogs to allow for optical+IR photometry for all WISE sources. 
The optical to WISE galaxy colors (e.g., $z$-$w1$) have strong redshift dependence and are therefore well suited for getting high-quality cluster redshifts. The $z$-$w1$ and $w1$-$w2$ color variation of passive galaxies with redshift is illustrated in Fig.~\ref{fig:colordep} using DES measurements in the COSMOS field.

One downside of WISE $w1$ is the large PSF ($\sim6"$), which becomes a problem in dense regions such as the cores of galaxy clusters. In such dense regions, the separation of individual sources and the deblending of fluxes is a challenge. Here the improved modeling of crowded regions in unWISE compared to previous ALLWISE catalog becomes relevant.

Finally, we account for masks or missing data in the different surveys by deriving separate richnesses for cluster regions with different coverage (DES + $w1$ only, WISE $w1$ only, WISE $w1$, $w2$ \& DES) and sum them for the final cluster richness estimate. With this approach we must track only the masked area in $w1$ imaging.
The total richness in the high-redshift code is therefore given as
\begin{align}
\label{eq:richnto}
 \lambda_{\mathrm{HZ}}(z)=&\lambda_{\mathrm{DES+w1}}(z)+\lambda_{\mathrm{DES+w1+w2}}(z)+\\&\lambda_{\mathrm{w1+w2}}(z)+\lambda_{\mathrm{w1}}(z),\nonumber
\end{align}
where the individual richnesses are defined in the same manner as the standard MCMF richness \citep[see][]{Klein18,Klein19}, with the color-weights depending on the availability of the bands ($i,z,w1,w2$ for $\lambda_{\mathrm{DES+w1+w2}}$, $i,z,w1$ for $\lambda_{\mathrm{DES+w1}}$, $w1$, $w2$ for $\lambda_{\mathrm{w1+w2}}$ and no color weight for $\lambda_{\mathrm{w1}}$).

The high-$z$ cluster confirmation code has been applied to all candidates and over a redshift range from $0.63<z<2.0$. Similar to the optically-based MCMF code, we perform runs along random lines-of-sight and calculate \fcont\ for potential counterparts to the SZE candidate. We make use of clusters with spectroscopic redshifts available for a subset of the sample and calibrate the WISE-based measurements. We further make use of the overlap in redshift between the optically-based MCMF and the high-z WISE-based run of $0.63<z<1.3$ to compare richnesses.

\begin{figure}
\includegraphics[keepaspectratio=true,width=0.97\columnwidth]{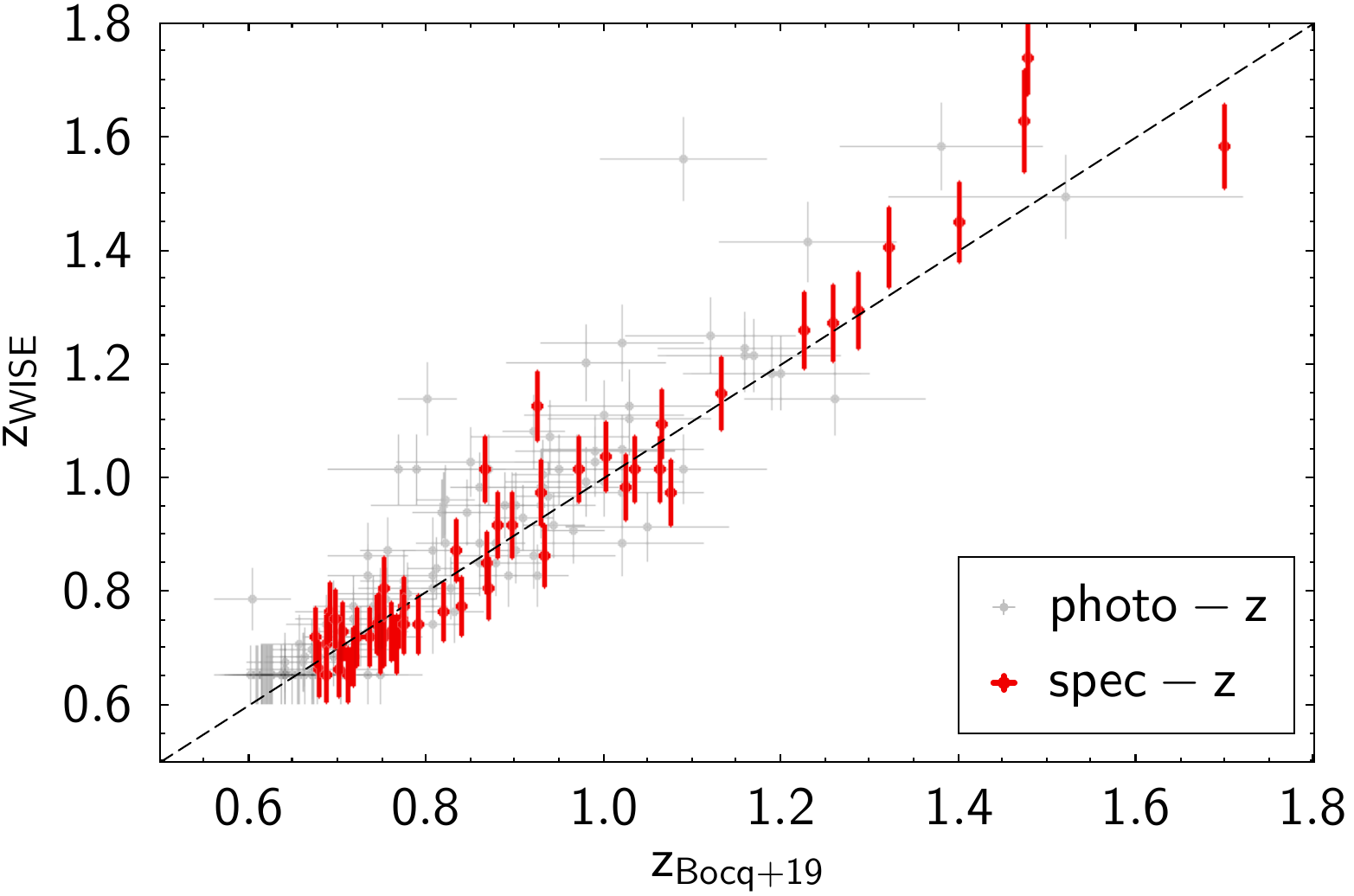}
\vskip-0.10in
\caption{Comparison of redshift estimates from previous SPT-SZ catalog \citep{Bocquet19} and those derived with the WISE-based high-z code. Spectroscopic redshifts are shown in red. Wise-based redshifts show generally good agreement with spectroscopic redshifts over the full redshift range although they are only used for clusters at $z>1$ in this work.}
\label{fig:highzperf}
\end{figure}

\begin{figure}
\includegraphics[keepaspectratio=true,width=0.97\columnwidth]{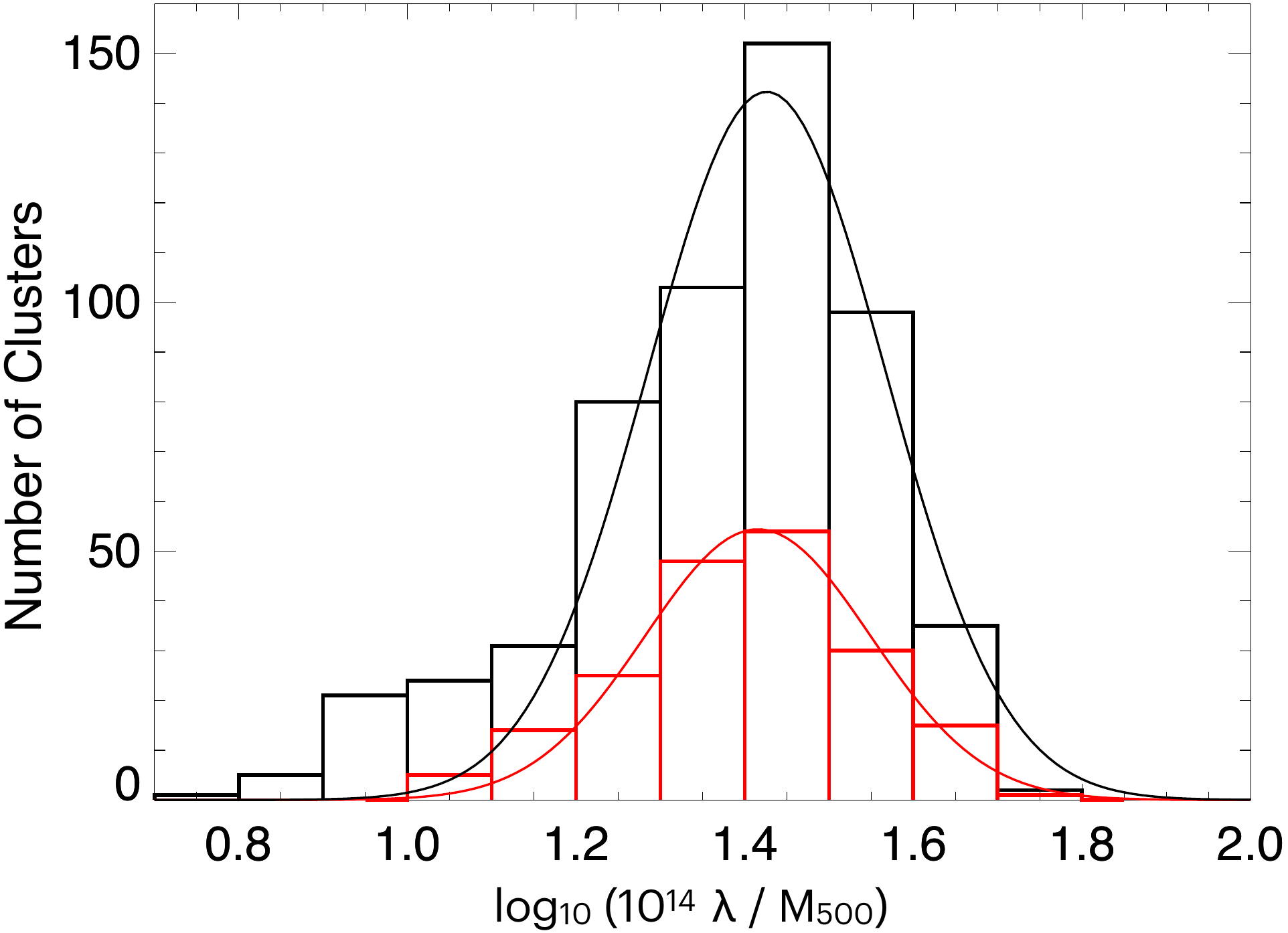}
\vskip-0.10in
\caption{Distribution of richness $\lambda$ over SZE-based mass estimate $M_{500}$ for richness measurements from the DES-only MCMF code (black) and the WISE-based high-z code (red). Continuous curves show the best fit normal distributions to values above log$(\lambda/M_{500})=1.2$ with best-fit standard deviations of $\sigma=0.14$ (black) and $\sigma=0.13$ (red).}
\label{fig:highzrichnessperf}
\end{figure}

 In Fig.~\ref{fig:highzperf} we show a comparison between the redshifts obtained with the high-z code and the redshifts provided for the previous catalog \citep{Bocquet19} for clusters with \fcont$<0.2$, showing reasonable agreement between WISE informed redshifts and the redshifts coming from dedicated optical, IR and spectroscopic follow-up. To avoid complicated modeling of the richness-observable and richness-mass relation it is further favourable that both richnesses share an approximately similar scatter.
For that reason we investigate the ratio of richness to the SZE-based mass estimate. Using the assumption that the richness-mass slope is approximately one and that there is no redshift evolution of the scaling relation, this ratio is a measure of the scatter of the lambda-mass relation. In Fig.~\ref{fig:highzrichnessperf} we show this ratio for the DES-only measurements of the $\xi>4.25$ subsample (see Table \ref{tab:sample}) and also for the high-z code measurements, here with the additional requirement that the cluster redshift must lie at $z>0.63$. As visible in Fig.~\ref{fig:highzrichnessperf}, the width and the mean of the distributions for the high-redshift code and the DES-only code appear very similar.  Both exhibit some deviation from a normal distribution. Fitting a normal distribution for the close-to-normal part of the distribution above log$(\lambda/M_{500})=1.2$ yields consistent mean ratios ($\mu_\mathrm{DES}=1.43\pm0.02$, $\mu_\mathrm{WISE}=1.42\pm0.01$) and standard deviations ($\sigma_\mathrm{DES}=0.14\pm0.02$, $\sigma_\mathrm{WISE}=0.13\pm0.01$), providing evidence that the two richness measurements exhibit similar relations to the SZE-based mass estimates. 
The cross-over in the ability to confirm cluster candidates is in the $1<z<1.3$ regime where the mid-IR selection of WISE starts to see more of the cluster population than is visible in the DES data.



\subsection{Optical estimators of cluster dynamical state}\label{sec:dynstate}
Following our previous work on X-ray selected clusters from ROSAT and eROSITA \citep{Klein19,Klein22} we provide for SPT-SZ MCMF clusters estimators related to cluster
morphology or dynamical state. Here we briefly describe the different estimators and refer the interested reader to our previous work for details \citep[see also][]{Wen13} .

We provide six dedicated measurements related to the morphological appearance of the galaxy cluster in the optical data. Additionally, the offsets between SZE and default optical centre
as well as SZE and a centre derived from fitting a 2D model to the galaxy distribution are presented and can be used as measures of the cluster dynamical state.
The 2D model centre is extracted while measuring the cluster morphology estimator $\delta$ \citep[described in][]{Klein19,Klein22,Wen13}.
The estimator $\delta$ measures the normalised deviation from a smooth two dimensional elliptical King model \citep{king62} fitted on the smoothed galaxy density map of red sequence cluster galaxies.
Besides the normalised deviation from the model, we also provide the
ellipticity $\epsilon$ of the fitted model as a measure of cluster morphology.
The ridge flatness $\beta$, compares the concentration of fitted one dimensional King profiles along different angular wedges and is the ratio of the lowest concentration value to the average concentration. Low-mass clusters falling into a massive cluster will cause the radial galaxy density profile to flatten towards the merger direction causing a low value of $\beta$. A third estimator  is the asymmetry factor $\alpha$ \citep{Wen13},
which measures the normalised average difference between pixel values in the galaxy density maps for pixels lying across from each other with respect to the cluster centre.
All four estimators are correlated with one another; they are based on the same galaxy density maps and associated noise, and they are sensitive to similar features-- primarily the asymmetry \citep{Klein22}. 

The last set of two estimators are derived by running SExtractor \citep{bertin96} on the passive galaxy density map. We use the resulting source list to identify nearby structures close to the main cluster and list the distance in terms of $R_{500}$ as well as the ratio of the flux\_auto measurement of the main and the second structure. The flux ratio can be thought of as a richness ratio, and it therefore serves as a proxy of the mass ratio between the two structures in question. The combination of both estimators makes it possible to select merger candidates
or cluster pairs that exhibit a certain mass ratio.


 \begin{figure}
\includegraphics[keepaspectratio=true,width=0.99\columnwidth]{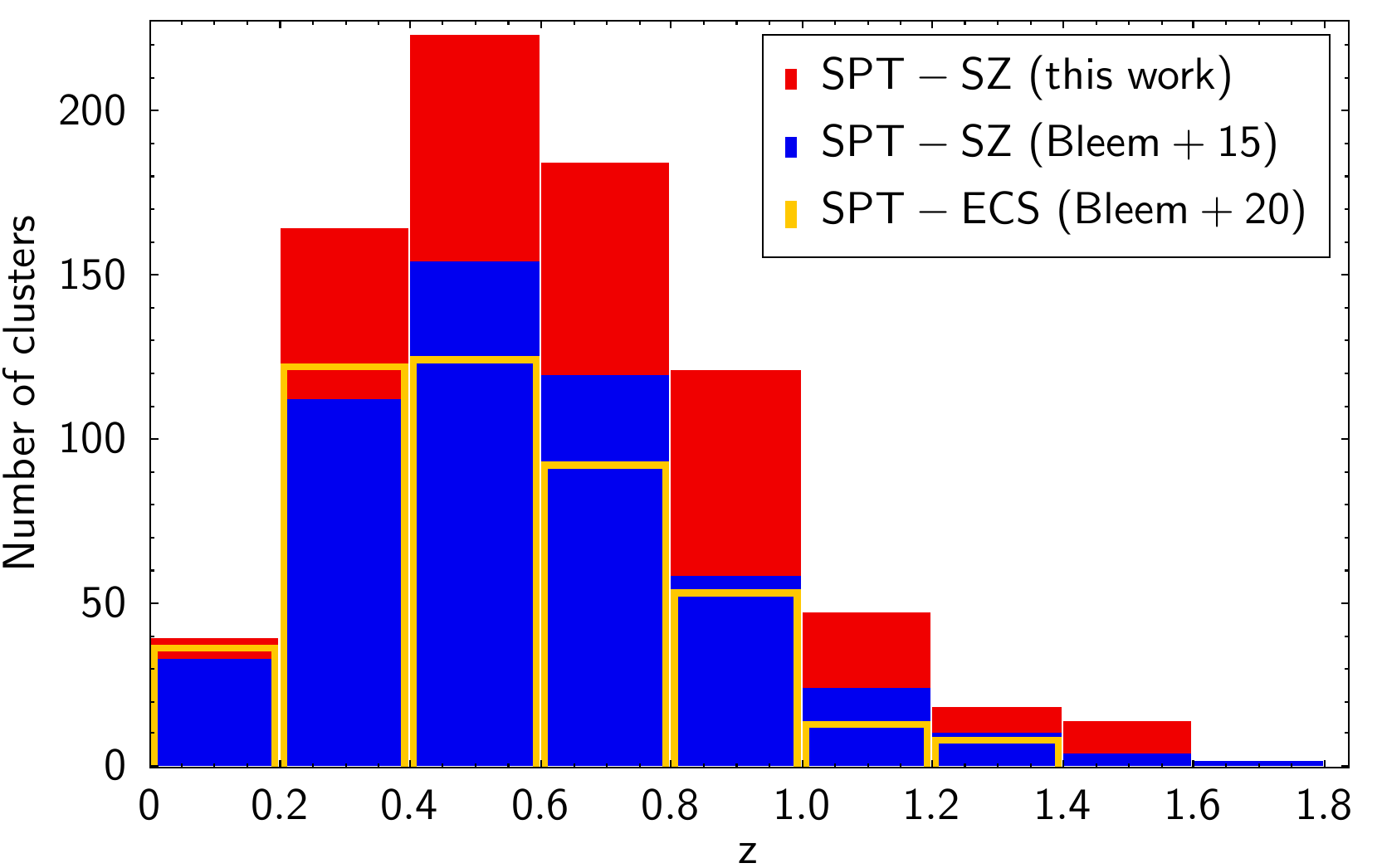}
\vskip-0.10in
\caption{Redshift distribution of the new SPT-SZ MCMF cluster sample containing 811 clusters with 9\% contamination (red background) in comparison to the previous SPT-SZ catalog (blue) and the SPTpol Extended Cluster Survey (SPT-ECS) catalog in yellow.}
\label{fig:redshiftdist}
\end{figure}

\begin{table}\caption{Properties of the SPT-SZ MCMF cluster catalog along with three subsamples. The table shows sample name, selection criteria $f_\mathrm{cont}^\mathrm{max}$ and $\xi_\mathrm{ min}$, the expected final sample purity, the completeness with respect to the SZE selection, the total number of confirmed clusters and those above redshift $z=0.25$.}
    \label{tab:sample}
    \centering
    \begin{tabular}{lcccccc}
        \hline
        \hline
        &  &  &  Purity & Comp. &  & $N_\mathrm{cl}$\\
        Sample & $f_{\mathrm{cont}}^{\mathrm{max}}$ & $\xi_\mathrm{min}$ & $[\%]$ & $[\%]$ & $N_\mathrm{cl}$ & (z>0.25)\\
        \hline
        SPT-SZ MCMF & 0.20 &  4.00  &  91.0 & 95.0 & 811 & 733\\
        $\xi>4.25$ & 0.125 & 4.25 & 96.0 & 96.5 & 640 & 581 \\
        $\xi>4.5$ & 0.13 & 4.50 & 98.0 & 96.5 & 527 & 480 \\
         \hline
    \end{tabular}
\end{table}

\section{cluster catalog}
\label{sec:SPT-SZ MCMF}

We present the new SPT-SZ MCMF cluster sample in Section~\ref{sec:sample} and then discuss the sample contamination (Section~\ref{sec:contamination}) and completeness (Section~\ref{sec:incompleteness}). In Section~\ref{sec:solidangle} we discuss the impact of DES masking on the survey solid angle. Finally, we present the results of the cluster morphological or dynamical state estimators in Section~\ref{sec:dynamical_state}.

\subsection{Defining the SPT-SZ MCMF galaxy cluster sample} 
\label{sec:sample}
As detailed in Section~\ref{sec:MCMF}, the MCMF \fcont\ measurements for each candidate provide a means of defining cluster samples with the desired contamination level.  For the catalog we present here, we adopt an \fcont\ threshold $f_\mathrm{cont}^\mathrm{max}=0.2$, which allows for 20\% of the original contamination present in the SPT-SZ candidate list to slip through into the confirmed cluster sample, which we call SPT-SZ MCMF.  As we will show in detail in Section~\ref{sec:contamination} we do have measurements of the amount of contamination of the original SPT-SZ candidate catalog as a function of SPT-SZ detection signal-to-noise $\xi$. For $\xi>4.0$ we measure an original contamination $f_\mathrm{SZE-cont}=45$\%.  
The final sample is therefore expected to have $0.2\times0.45=9$\% contamination and it contains 811 clusters.
The \fcont\ selection threshold introduces incompleteness in the catalog at the level of 5\%, because while the \fcont\ selection filters out contaminants it also removes some real, low-richness clusters.  Details of this sample and other subsamples described below are shown in Table~\ref{tab:sample}, which lists the \fcont\ threshold, SZE S/N threshold, purity, completeness, number of confirmed clusters and number of clusters at $z>0.25$.  Here the listing of clusters above $z=0.25$ is of special relevance for cosmological analyses, which have typically excluded lower redshift systems due to the angular filtering in the SPT cluster selection \citep[see, e.g.,][]{dehaan16,Bocquet19}.

In selecting the best suited cluster sample for a given science investigation, different sample criteria can be more or less important. To guide the reader, we provide here two additional subsamples of the SPT-SZ MCMF cluster catalog by varying $f_\mathrm{cont}^\mathrm{max}$ and $\xi$ selections. 

The first of the two subsamples ($\xi>4.25$ in Table~\ref{tab:sample}) has SZE selection thresholds $\xi>4.25$ and $f_\mathrm{cont}^\mathrm{max}=0.125$. Our current understanding is that cluster subsamples with $\xi>4.25$ are better suited for studies relying on well behaved SZE-based cluster masses (e.g., mass-observable scaling relations or cluster counts cosmology analyses). Furthermore, when modeling cluster counts to derive cosmological constraints, it simplifies the analysis if the subsample contamination is low enough that it does not require detailed modeling. A guideline here is that the contamination fraction is at or below the level of the Poisson noise associated with the full subsample. Given the sample sizes here, the target upper limit for the contamination ranges between 3\% and 4\%, and this is met for both of the subsamples presented. A similar target can be set for the completeness of the sample, relative to its original SZE selection. As we will show in Section~\ref{sec:incompleteness}, the particular choice of $f_\mathrm{cont}^\mathrm{max}$ used for this subsample meets requirements for both, purity and completeness. We also note that the incompleteness due to optical cleaning can be accounted for (see e.g. \citet{Grandis20,Chiu23}).

The $\xi>4.5$ subsample represents a more conservative selection in $\xi$ and a looser cut in $f_\mathrm{cont}^\mathrm{max}=0.13$, which remains in the well-tested $\xi$-regime. The contamination is low in the SZE candidate list, the predicted false detections from simulations and observations are in good agreement. The optical selection plays an insignificant role, introducing 3.5\% incompleteness through \fcont\ selection while providing a high (98\%) purity sample. The $\xi>4.5$ subsample will further be part of the upcoming SPT cosmological analysis, which includes modeling of the impact optical cleaning on the sample selection.

The redshift distribution of the SPT-SZ MCMF cluster sample is shown in Fig.~\ref{fig:redshiftdist}, where it is compared to the previously released SPT-SZ catalog \citep{Bleem15} with updated redshifts from \citet{Bocquet19} and the SPTpol Extended Cluster Survey catalog \citep[SPTpol-ECS][]{Bleem20}. Of the 811 confirmed clusters in the SPT-SZ MCMF catalog, 91 are at $z>1$. This is a substantial increase compared to the 516 clusters in the previous catalog, and it more than doubles the number of high-redshift clusters. The DES data cover only 92\% of the SPT-SZ sources, and consequently we miss 34 confirmed clusters from \citet{Bleem15}, and would expect $\sim$69 clusters to lie outside the footprint.
Adding the other published SPT-based SZE cluster catalogs, SPTpol-ECS \citep{Bleem20} and SPTpol 100d \citep{SPT100d}, and excluding duplicates, we obtain a combined catalog containing 1,343 clusters, exceeding the number of confirmed SZE clusters from the second Planck catalog of Sunyaev-Zeldovich sources \citep{PSZ2}, but lying below the number of S/N>4 candidates presented by the ACT collaboration \citep{ACTDR5}.

 \begin{figure}
\includegraphics[keepaspectratio=true,width=0.99\columnwidth]{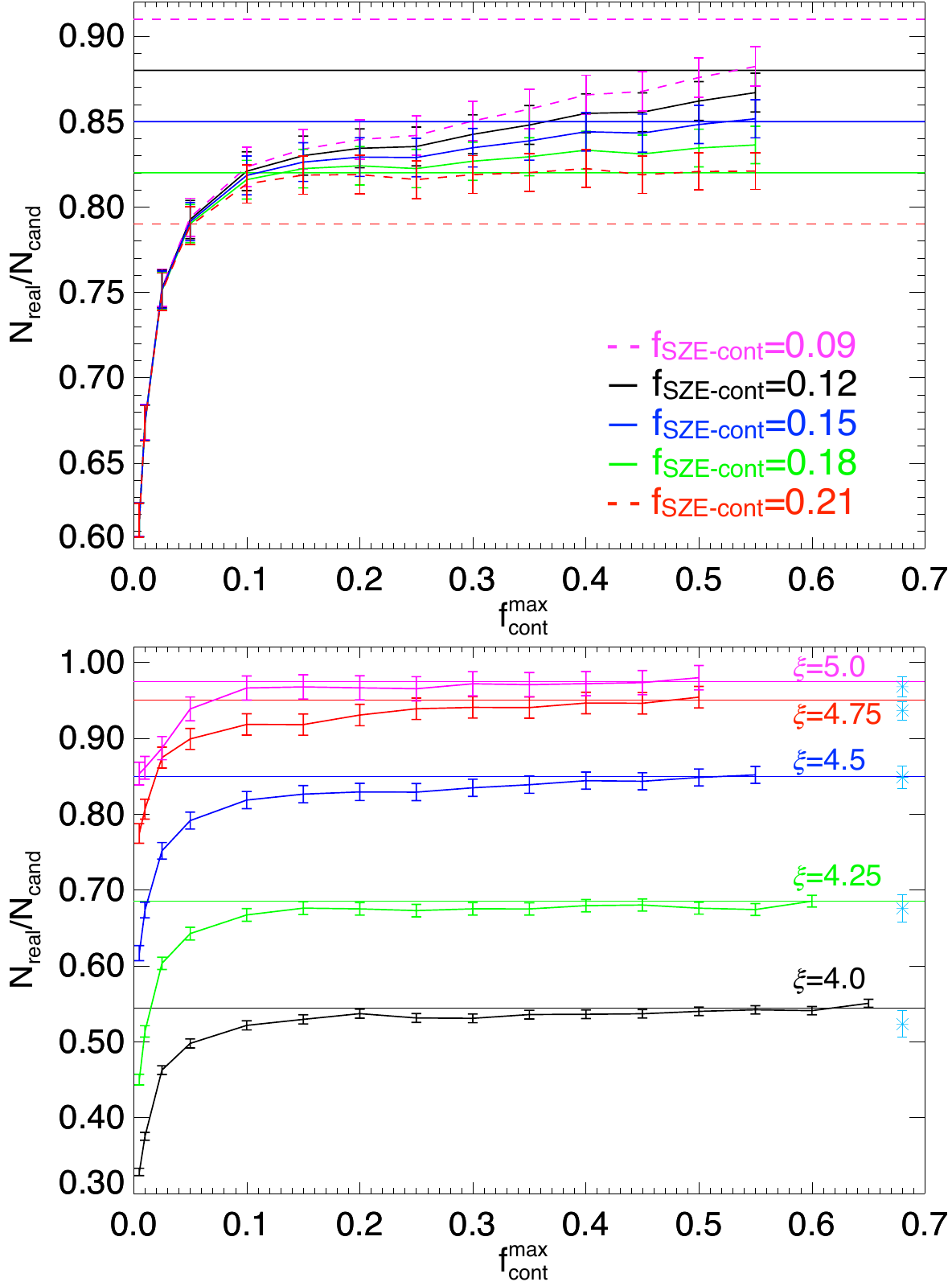}
\caption{Empirical estimation of the initial contamination $f_\mathrm{SZE-cont}$ in different SPT-SZ subsamples.
Top: Example of the $f_\mathrm{cont}^\mathrm{max}$-based method (see discussion in Section~\ref{sec:contamination}) applied to the $\xi>4.5$ SZE-selected sample with different assumptions for the initial contamination.  
The best fit initial contamination of 15\% is shown in blue.
Bottom: Best fit results for five different SPT-SZ selection thresholds $\xi$=5.0, 4.75, 4.5, 4.25 and 4.0 arranged from top to bottom that indicate an initial purity of 97.5\%, 95\%, 85\%, 69\% and 55\%, respectively.  For each case the cyan point at \fcont=0.8 shows an independent purity estimate from the mixture model method using the distribution of candidates in $\logTen(10^{14} \lambda/ M_{500})$. Both methods are in good agreement with eachother for all thresholds in $\xi$.}
\label{fig:puritycomp1}
\end{figure}

 \begin{figure}
 \includegraphics[keepaspectratio=true,width=0.99\columnwidth]{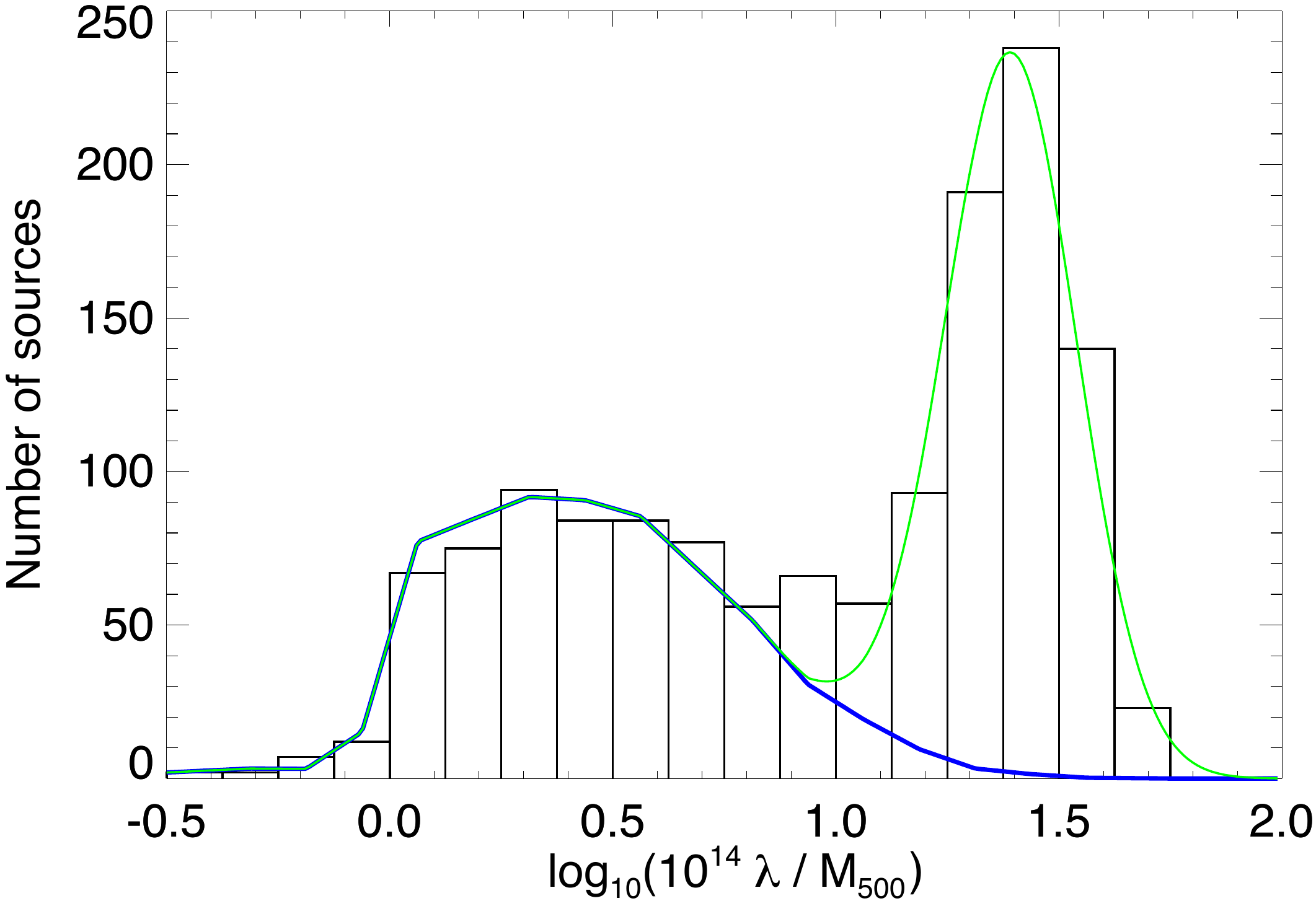}
\caption{Example of the empirical estimation of the initial contamination $f_\mathrm{SZE-cont}$ based on the distribution of candidates in $\logTen(10^{14} \lambda/ M_{500})$. The model (green) of the richness distribution of SPT-SZ candidates as a mixture of contaminants (in blue) and clusters. The composite model contains clusters modeled as a Gaussian distribution. The contaminant population is defined using measurements along random lines-of-sight.}
\label{fig:contaest2}
\end{figure}

\subsection{Initial contamination of SPT-SZ candidate lists}
\label{sec:contamination}

As discussed in Section~\ref{sec:MCMF}, the expected contamination fraction of a sample selected using a particular \fcont\ threshold (i.e.,  clusters with $f_\mathrm{cont}<f^{\mathrm{max}}_{\mathrm{cont}}$) is $f^{\mathrm{max}}_{\mathrm{cont}}\times f_\mathrm{SZE-cont}$.  Therefore, it is crucial to know the contamination fraction in the initial candidate list.  We use two methods to estimate $f_\mathrm{SZE-cont}$ for the different SPT-SZ candidate lists (i.e., with different SZE selection thresholds in $\xi$). The first follows our previous work in \citep{Hernandez22} and uses the fact that in \fcont$<f^{\mathrm{max}}_{\mathrm{cont}}$ selected samples, the completeness should reach 100\% for high values of $f^{\mathrm{max}}_{\mathrm{cont}}\sim1$.
The number of expected real clusters is
\begin{equation}\label{eq:nreal_nclus}
N_{\mathrm{real}} (f^{\mathrm{max}}_{\mathrm{cont}} ) = N_{\mathrm{MCMF}}(f_{\mathrm{cont}} < f^{\mathrm{max}}_{\mathrm{cont}} ) \left[1 - f^{\mathrm{max}}_{\mathrm{cont}}f_\mathrm{SZE-cont}\right],
\end{equation}
where $N_{\mathrm{MCMF}}(f_{\mathrm{cont}} < f^{\mathrm{max}}_{\mathrm{cont}})$ is the number of systems in the MCMF confirmed catalog with \fcont\ values below $f^{\mathrm{max}}_{\mathrm{cont}}$.
The ratio $N_{\mathrm{real}}/N_\mathrm{cand}$, where $N_\mathrm{cand}$ is the number of SPT-SZ candidates, should reach but not exceed the expected purity of the candidate list 
$(1-f_\mathrm{SZE-cont})$. 
Incorrectly estimating $f_\mathrm{SZE-cont}$ would lead to inconsistencies, such as finding (1) more real clusters than allowed or (2) falling numbers of real clusters at high $f^{\mathrm{max}}_{\mathrm{cont}}$.

One illustrative example for an SPT-SZ sub-sample with $\xi>4.5$ is shown in the top panel of Fig.~\ref{fig:puritycomp1}. Here we show the behaviour of $N_{\mathrm{real}}/N_\mathrm{cand}$ for five different values of initial contamination fractions from 9 to 21\% in steps of 3\%. The horizontal lines show the expected purity $(1-f_\mathrm{SZE-cont})$ for the curves with the same color. As can be seen, setting the initial contamination $f_\mathrm{SZE-cont}$ too high (lowest two curves in red and green) causes an over prediction of real clusters (lines with data points) relative to that expected number given the assumed contamination level (flat line of same color).  This clear inconsistency excludes these high contamination levels.

For the lower assumed contamination cases with $f_\mathrm{SZE-cont} \le 0.12$, the curves with data points (black and magenta) continue rising over the full range of \fcont. This is a very unlikely scenario, given the expected richnesses of SPT clusters ($\lambda>20$) and the richness levels probed at \fcont$>0.6$ ($\lambda\approx 2$). For the initial contamination level of $f_\mathrm{SZE-cont}$=0.15 (black curve with data points) we find a stable solution where the fraction of real clusters converges to the expected contamination fraction and then remains roughly constant above \fcont>0.4.  This is a clear indication that this $\xi>4.5$ SPT-SZ candidate list has $\approx$15\% contamination.

In the lower panel of Fig.~\ref{fig:puritycomp1} we show the results for five SPT-SZ candidate lists with different thresholds in signal-to-noise $\xi$ of 5, 4.75, 4.5, 4.25 and 4. In these cases we remeasure \fcont\ for each subsample using the appropriate signal-to-noise thresholds in the candidate and random sample.  One can read off the purity of these samples to be 97.5\% (magenta), 95\% (red), 85\% (blue), 70\% (green) and 55\% (black), respectively.  As expected, going to lower SPT-SZ signal-to-noise decreases the purity of the initial candidate lists.  But as we have previously emphasized, the MCMF algorithm enables the removal of a large fraction of the contamination and the delivery of an overall larger cluster sample.  This enlarged sample extends to lower masses at all redshifts and therefore typically also extends to higher redshift.

The second and main method to derive the level of initial contamination for different SPT-SZ candidate lists makes use of the richness distributions of the candidates and along random lines-of-sight and follows our previous work on X-ray selected clusters \citep{Klein22,Klein23}. Contrary to the first method it does not rely on \fcont\ selection or on the correct derivation of \fcont. To estimate the initial contamination, we model the distribution of candidates in $\logTen(10^{14} \lambda/ M_{500})$ space as a mixture of a contamination and a cluster model. The contamination model is directly derived from the measurements along random lines-of-sight as a histogram in $\logTen(10^{14} \lambda/ M_{500})$ that can be re-scaled to adopt for different amounts of contamination. As richness scales approximately linear with mass, the cluster population in $\logTen(10^{14} \lambda/ M_{500}$) can simply be assumed to be normally distributed. The total model therefore consists of just four parameters, three for the normal distribution and just one, the normalization, for the contaminant distribution. As example, the observed and the fitted model for the $\xi>4$ candidate list is shown in Fig.~\ref{fig:contaest2}.
For the estimate of the contamination we are solely interested in the best-fitting contamination model, which is predominantly determined at $\logTen(10^{14} \lambda/ M_{500})<1$, where the cluster component plays no significant role.

The resulting purity estimates of this sample together with results for the other candidate lists with different $\xi$ thresholds are shown as cyan points in the lower panel of Fig.~\ref{fig:puritycomp1}.
The results of both methods are consistent.
These two methods can be used to derive the level of initial contamination in a candidate list even in the case where contaminants and clusters are not clearly separated in, e.g., a space of $\lambda$ versus redshift.  

\subsection{SPT-SZ MCMF incompleteness due to optical cleaning}
\label{sec:incompleteness}

As already mentioned, the MCMF algorithm for excluding contamination can also exclude real clusters.
The impact of the \fcont-based selection, which is essentially a redshift dependent threshold in richness, can be modeled using the richness--mass relation \citep[e.g., see][]{Klein22}. In addition, a rough estimate of the overall completeness can be derived using the previously estimated initial contamination. The initial contamination defines the number of real clusters in the candidate list as well as the number of real clusters given the \fcont\ selection threshold. The differences between these two reflects the impact on the completeness of the real cluster sample.  The impact of \fcont-based cleaning is already clearly visible in Fig.~\ref{fig:puritycomp1} (see bottom panel) as the difference between the horizontal lines that show the fraction of the candidates that are real clusters $(1-f_\mathrm{SZE-cont})$ and the curves with data points that show the recovered fraction of real clusters as a function of the \fcont\ threshold employed.

\begin{figure}
 \includegraphics[keepaspectratio=true,width=0.99\columnwidth]{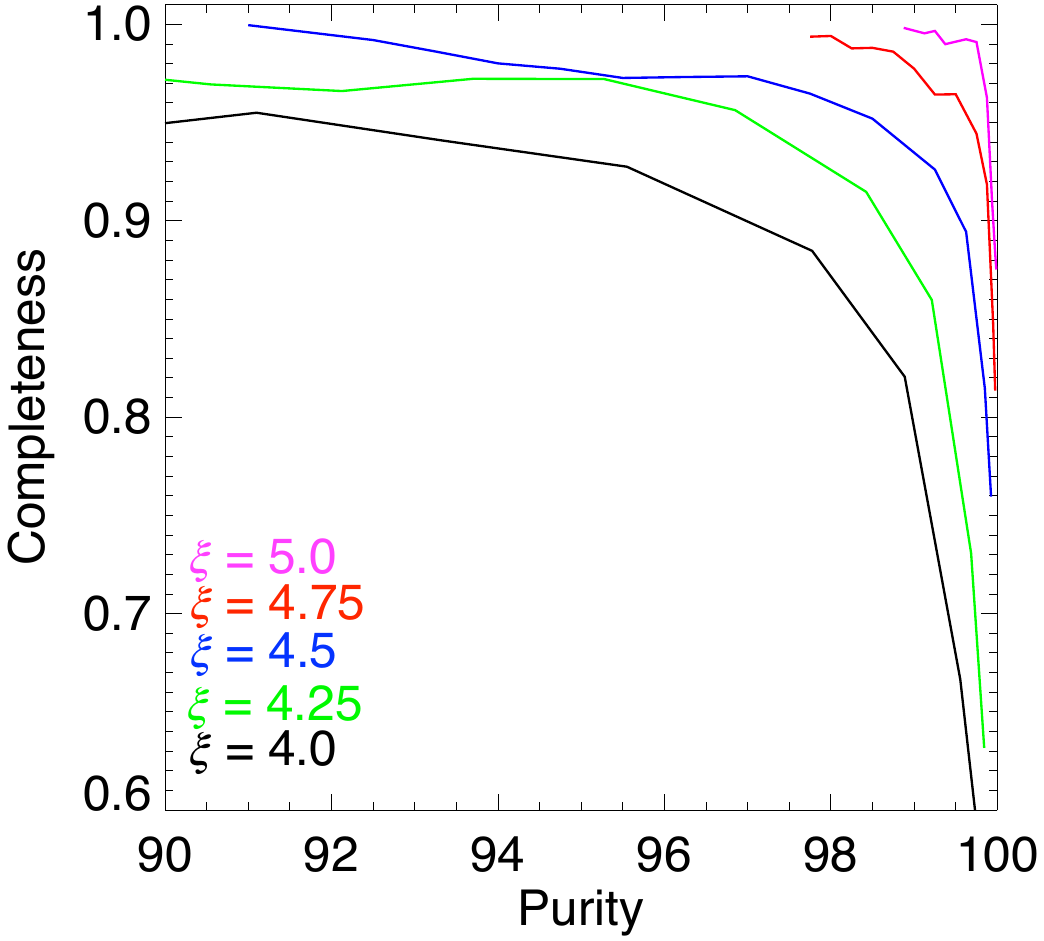}
\caption{The purity as a function of completeness is shown for five different thresholds in the SPT-SZ selection threshold $\xi$.  These curves are built for each sample by varying the MCMF defined optical selection \fcont\ threshold. Through tuning the \fcont\ threshold to lower values, one can create a final catalog with a increased purity at the cost of introducing additional incompleteness.}
\label{fig:complete}
\end{figure}

In Fig.~\ref{fig:complete} we show the expected completeness of the \fcont\ selected sample with respect to the number of real clusters in the SZE selected sample versus the purity;  this is shown for the same five SPT-SZ $\xi$ thresholds examined previously in Fig.~\ref{fig:puritycomp1}. Each curve is built by calculating the purity and completeness for a range of \fcont\ selection thresholds increasing from right to left.  The purity is derived as $1-f_\mathrm{SZE-cont}\times f^{\mathrm{max}}_{\mathrm{cont}}$ and the completeness is the fraction of real clusters that survive the \fcont\ selection.
 As one can see, the impact of the optical cleaning on the completeness remains mild ($\leq 5\%$) for all SPT-SZ subsamples until one reaches a purity of 95\% or above, after which the completeness drops precipitously.  Moreover, the highest purity samples tend to be smaller.  One must consider these impacts when selecting a sample for scientific analysis.
 
As discussed in Section~\ref{sec:sample} the expected amount of sample Poisson noise (i.e., important for cluster count statistics) is simply $1/\sqrt{N_\mathrm{clust}}$, and corresponds to 3.5-5\% for $\xi$ thresholds of 4 to 5 in the SPT-SZ sample.  This sets an upper limit  for the target contamination such that contamination need not be explicitly modeled.
While this translates into a completeness of 91\% for the $\xi>4$ sample, it results in >96\% completeness for higher $\xi$ threshold SPT-SZ subsamples, bringing the incompleteness below the level of Poisson noise of these sub-samples. This means for sub-samples with $\xi$ thresholds of 4.25 or higher we are able to construct cluster samples where contamination and incompleteness are both below the expected Poisson noise, which implies that these effects will have a sub-dominant impact.
Even in the contrary case, the sample incompleteness can be straightforwardly accounted for by modeling the selection in $\xi$ jointly with the requirement that the richness $\lambda$ exceeds the threshold corresponding to \fcont. A sample selected according to \emph{both} variables $\xi$ and $\lambda$ is then complete with respect to a model that accounts for the joint selection.
This modelling approach has already been successfully applied in the cosmological analysis of a real cluster sample \citep{Chiu23}.
However, currently, explicit modelling of contaminants in ICM-selected samples is still lacking. This explains the choice of a very clean (98\% pure)
selection for the $\xi>4.5$ sub-sample used in the upcoming SPT-based cosmological study (Bocquet et al., in prep.). This study explicitly includes modelling of the incompleteness due to MCMF-based cleaning but relies on high purity to avoid modelling contamination.

 \begin{figure*}
\includegraphics[keepaspectratio=true,width=0.32\linewidth]{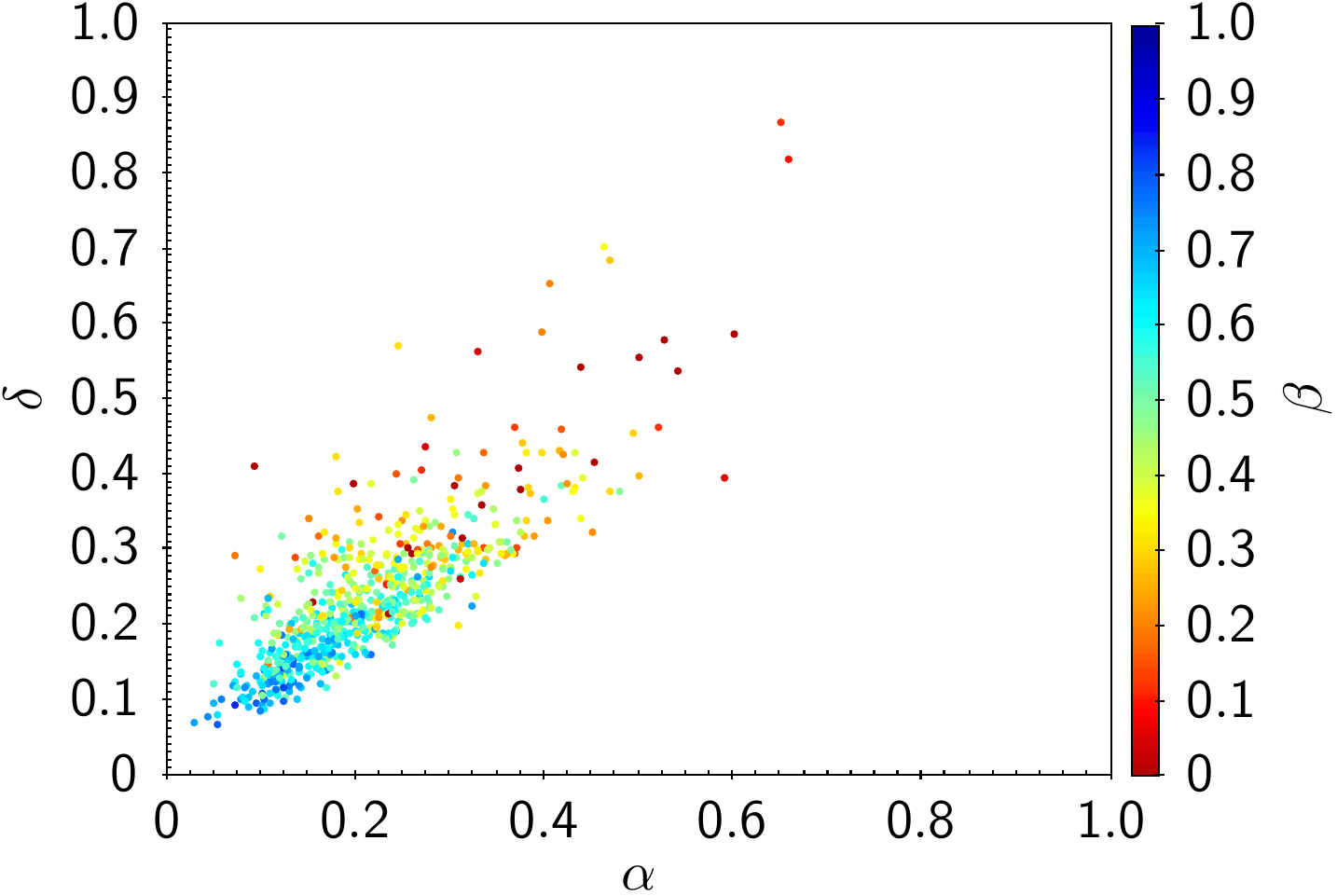}
\includegraphics[keepaspectratio=true,width=0.32\linewidth]{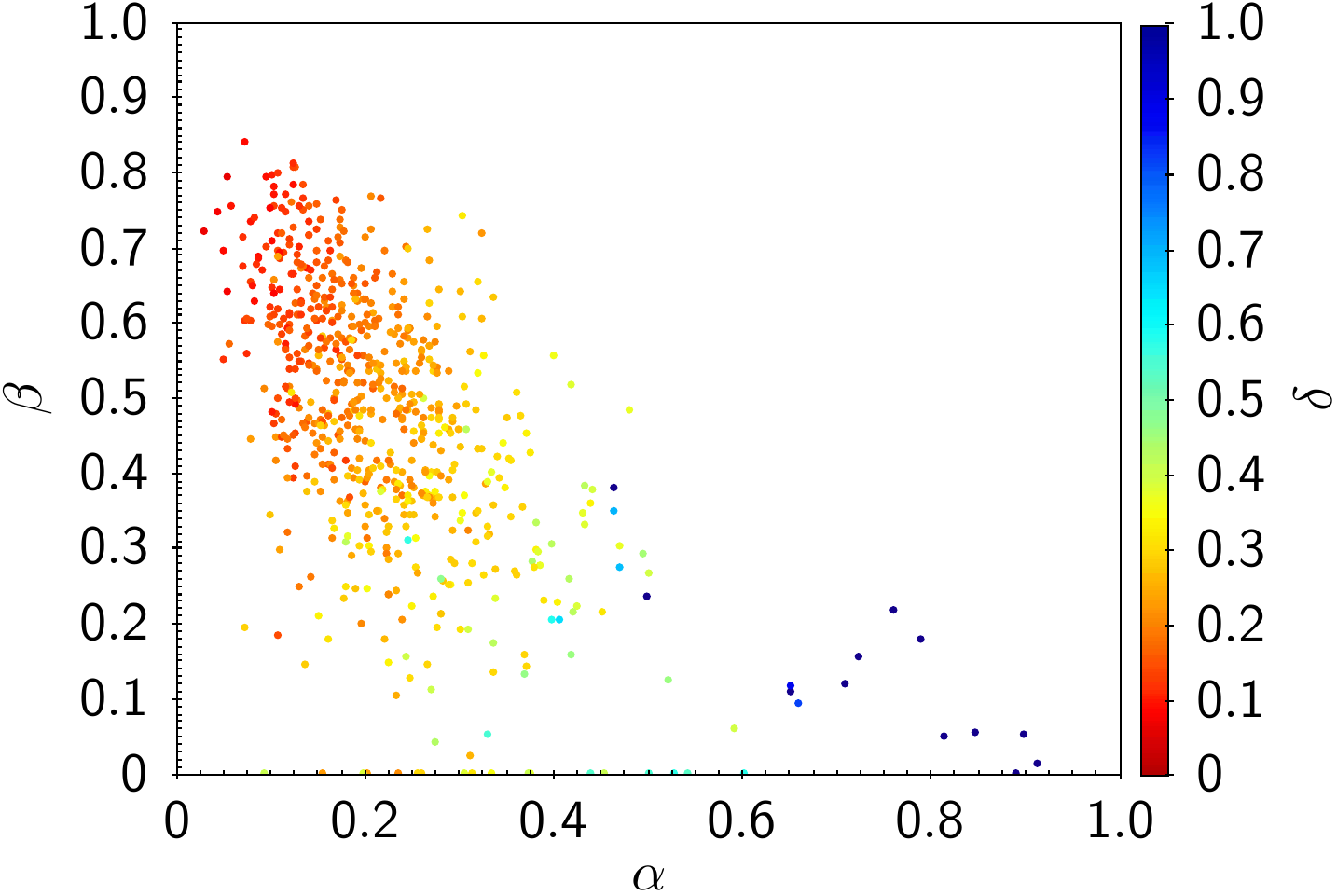}
\includegraphics[keepaspectratio=true,width=0.32\linewidth]{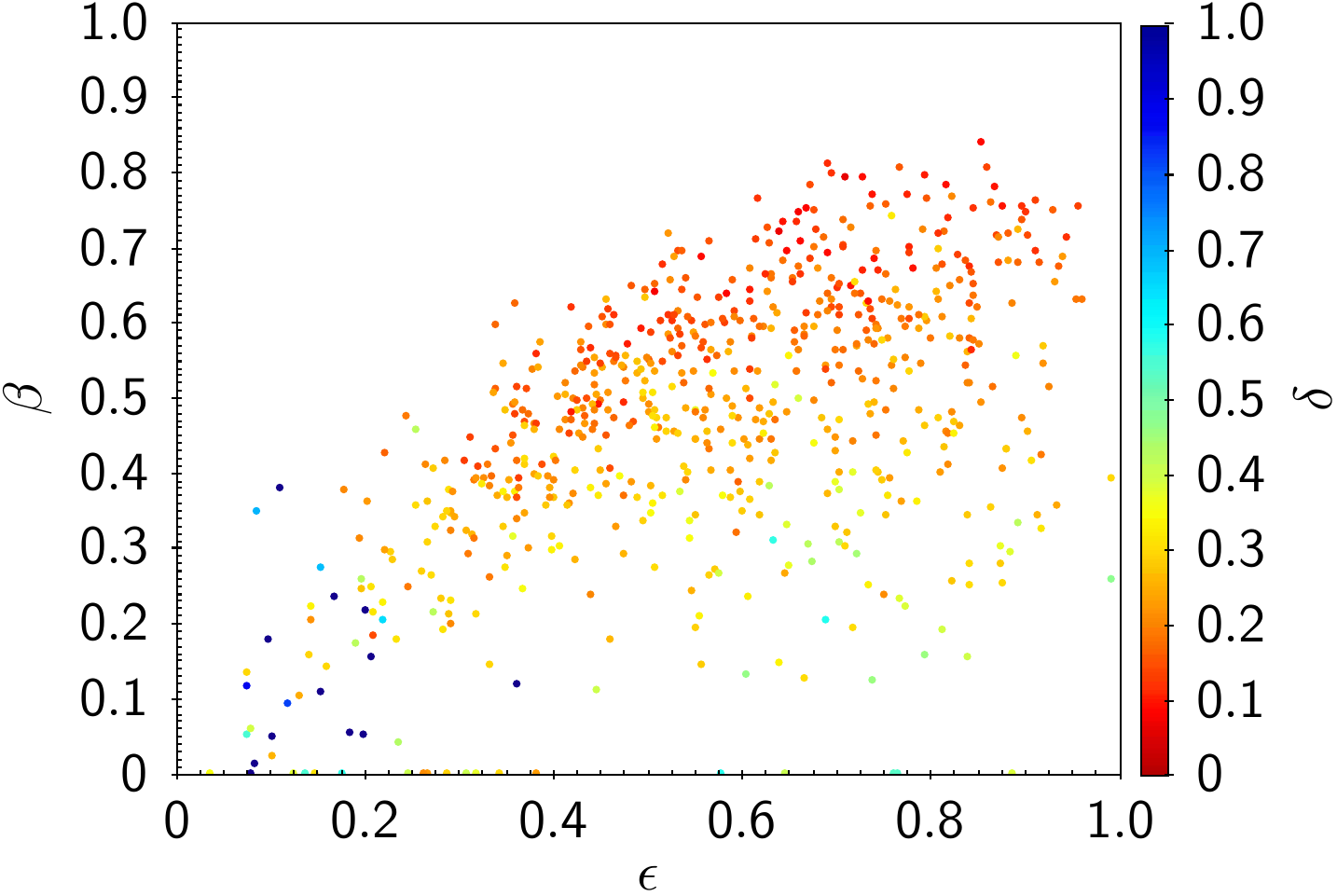}
\vskip-0.10in
\caption{Comparison of different optical morphology estimators described in Section~\ref{sec:dynamical_state} for SPT-SZ MCMF clusters (\fcont<0.2) with $\lambda>25$. Estimators probe different merging properties, but are well correlated.}
\label{fig:dynestis}
\end{figure*}

\subsection{Impact of DES masking on SPT-SZ MCMF survey solid angle}
\label{sec:solidangle}
The previous section covers the impact of optical cluster confirmation on the completeness and purity of the final cluster catalog within the general DES footprint. One additional problem that arises with optical confirmation is that within the DES footprint there are areas with missing optical data due to, e.g., bright stars or a lack of data due to poorly performing CCDs or even chip gaps.  We follow these regions by building a sky mask for the optical data.
The impact of missing data on cluster confirmation depends on the location and size of the masked region with respect of the SPT-SZ candidate position and effective size $\theta_{500}(z)$.  Prior to confirmation the redshift and therefore the corresponding cluster size $\theta_{500}(z)$ is not well known; therefore, we use the sky masking fraction within a fixed angular distance around the candidate locations to characterize the importance of masking.  

To estimate the impact of masking on cluster confirmation, we use the confirmation fraction as a function of the masking fraction. We explore three different apertures sizes with radii of one, two and three arcminutes to test the sensitivity to masking. As a baseline we use clusters not showing masking and derive confirmation fractions $N($\fcont$<0.2)/N_\mathrm{cand}$ of $0.588\pm0.014$, $0.593\pm0.014$ and $0.595\pm0.015$ for the three apertures in discussion. 
Looking into the confirmation fractions we see that we would not have confirmed any candidate with mask fractions greater 0.56 in the two or three arcminute apertures and in total only one out of eight candidates with mask fractions greater 0.5. 
We therefore decided to re-define the minimum definition of a source to be considered within the DES footprint to have at least one source within 1 arcminute and a mask fraction within two arcminutes below 0.5, effectively reducing the footprint by 0.6\%.
There is no statistically significant impact visible on the confirmation fraction between mask fractions zero and 0.5. Taking all sources in that masking range we find confirmation fractions of  $0.52\pm0.05$, $0.52\pm0.04$ and $0.54\pm0.03$, consistent within two sigma from the baseline confirmation fractions. This residual effect can generally be accounted for by an overall re-scaling of the footprint area by 1.5\%. But we note that this correction is on the level of one sigma, given the uncertainty on the confirmation fraction of the unmasked clusters, this correction is likely not necessary for most studies using this sample.


\subsection{Optical morphology and dynamical state}
\label{sec:dynamical_state}
The morphology of a cluster can be an indicator of dynamical state, and so in principle the cluster morphology can be used to identify samples of clusters for the study of the dynamical evolution of clusters and cluster components.
With the SPT-SZ MCMF catalog we include optical morphological estimators of dynamical state for all confirmed clusters;  however, the quality of the measurements depends on richness and redshift.
Increasing the richness selection threshold will further improve the robustness of the estimates.  We therefore recommend restricting morphology analyses to the redshift range of $0.1<z<0.9$ and a richness $\lambda>40$. 
In Fig.~\ref{fig:dynestis} we present comparisons among the four morphology estimators $\alpha$, $\delta$, $\beta$ and $\epsilon$ that are described in Section~\ref{sec:dynstate}.  
As expected, the estimators are strongly correlated. 

\begin{figure}
\includegraphics[keepaspectratio=true,width=1\linewidth]{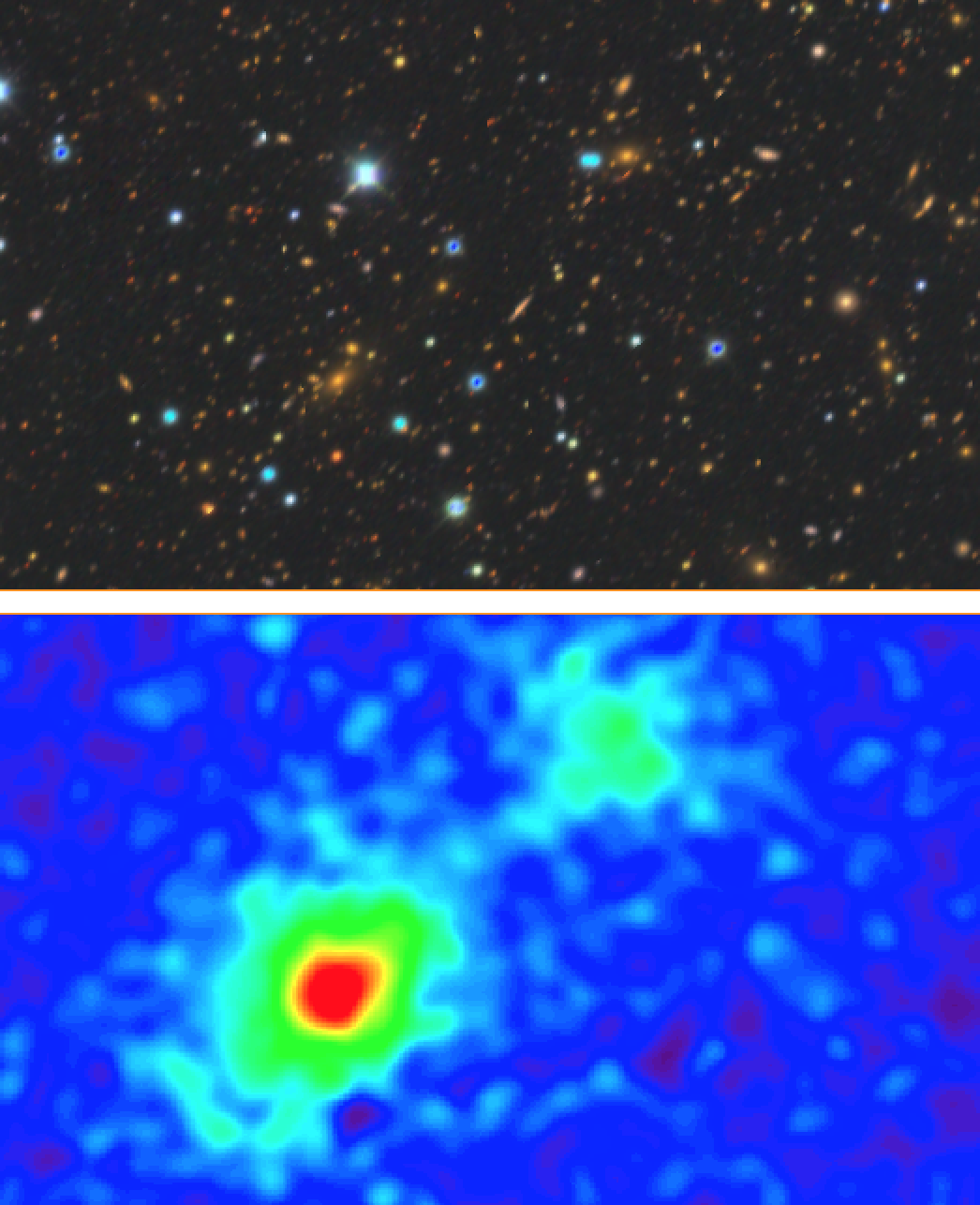}
\vskip-0.10in
\caption{Top: DES RGB-image of 9'$\times$5.5' region around SPT-CL~J0522--4818. Bottom: Smooth Chandra X-ray image of the same region. The cluster is classified as one of the most unrelaxed systems in optical while having a low X-ray-based disturbance estimate. The two clusters are likely in a pre-merger or early merger state, where the X-ray surface brightness distribution of the main system probed by the X-ray estimators is not yet affected by the merger process.}
\label{fig:SPT0522}
\end{figure}

A preliminary comparison to X-ray morphological merger estimators that trigger on the skewness and ellipticity of the ICM distribution \citep[e.g.,][]{mohr93,nurgaliev13,nurgaliev17} for a subset of clusters that have Chandra or XMM-Newton observations shows little correlation, underscoring that optical and X-ray merger indicators are sensitive to different stages of cluster mergers and are also affected differently by projection effects. A simple example for such a case of mismatching classifications is SPT-CL~J0522--4818, shown in Fig.~\ref{fig:SPT0522}. Therefore, we expect that these optical morphological estimators could be useful in combination with the established X-ray techniques for the purpose of creating a sequence of clusters covering a broad range of dynamical state.

\section{Catalog Validation}
\label{sec:validation}

 We validate SPT-SZ MCMF through comparison to several other catalogs in Section~\ref{sec:comparison}, carry out an examination of the contaminant distribution of the SPT-SZ candidate list in Section~\ref{sec:contaminants} and then carry out a modeling validation in Section~\ref{sec:counts} that employs parameter constraints from a cosmological analysis of the previous SPT-SZ sample.


\subsection{Comparison of SPT-SZ MCMF to other cluster catalogs}
\label{sec:comparison}

We compare the new catalog to three previously published SZE selected cluster catalogs.

 \begin{figure}
\includegraphics[keepaspectratio=true,width=0.97\columnwidth]{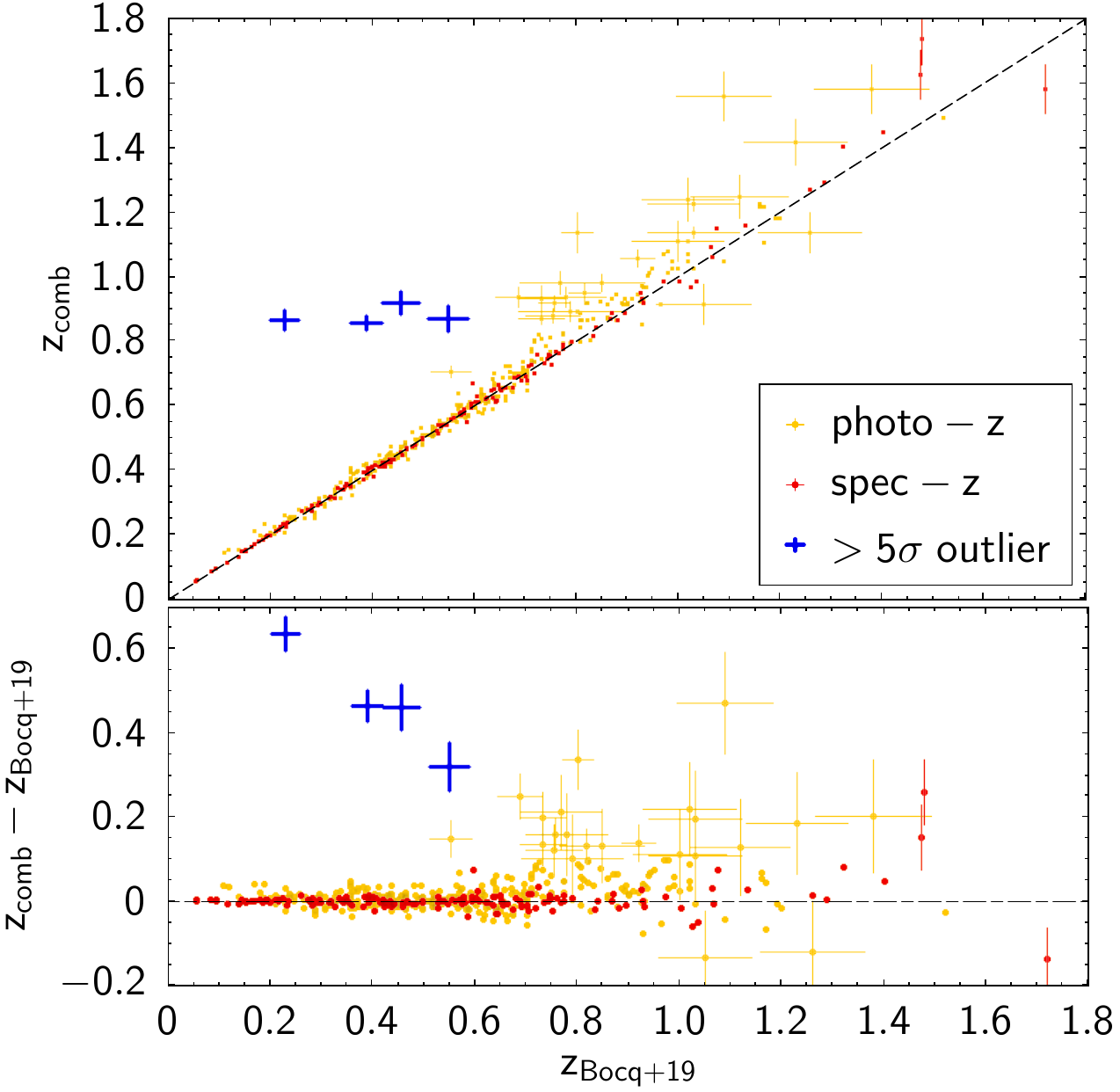}
\vskip-0.10in
\caption{Comparison between redshifts given in in this work and those presented \citet{Bocquet19}, split in photometric (yellow) and spectroscopic redshifts (red). The most significant outliers ($\Delta z > 5\sigma$) are shown in blue. All of those have a second ranked optical counterpart with redshifts in agreement with \citet{Bocquet19}. There is further a indication for a redshift bias in previous photo-z above $z=0.7$. For sake of readability we only show uncertainties for sources with $|\Delta z|>0.1$.}
\label{fig:photozcomp}
\end{figure}

\subsubsection{Previous SPT-SZ catalog}
To check for consistency we compare our results to the previous release of the SPT-SZ catalog \citep{Bleem15} and considering the updated redshifts provided in \citet{Bocquet19}.  The expected contamination of the SPT-SZ candidate list at $\xi>4.5$ adopted in the previous study is 15\%, and therefore the \fcont\ threshold 0.2 would correspond to an expected contamination in the final catalog of 3\%.
We find 481 clusters that have redshifts in both catalogs, and all but 4 have \fcont$<0.2$.  In all cases the previously published redshift estimate is consistent with the redshift presented here. 

The number of (4) unconfirmed systems, corresponding to $\sim$1\% of the previously confirmed $\xi>4.5$ sample, is consistent with the expected incompleteness due to MCMF based \fcont\ selection of 2\% given in Fig.~\ref{fig:complete}. Furthermore, some of the previously confirmed clusters could indeed be chance superpositions.
We fail to confirm SPT-CL~J0334--4645, the highest-redshift SPT cluster at $z=1.7$ and one of the lowest-redshift clusters SPT-CL~J2313--4243 at $z=0.056$. The latter is well identified in MCMF, but its richness is too low to meet the \fcont$<0.2$ selection. In addition, we fail to confirm SPT-CL~J0002--5557, which is listed to be at $z=1.15$, whereas our high-z analysis places this cluster at $z=1.37$ with an \fcont\ estimate of 0.45. The lack of red galaxies visible in the DES image indicates that this cluster needs to be beyond the MCMF DES redshift reach of $z\sim1.3$. In WISE the cluster is visible as one compact red blob, which may be the reason for the relatively high \fcont\ because counting cluster members for this compact cluster might have failed.  The last cluster, SPT-CL~J2005--5635, at $z=0.2$ shows a low richness resulting in \fcont$=0.31$. While the other three clusters do have matches in SZE or X-ray surveys, this cluster does not.
 
In Fig.~\ref{fig:photozcomp} we show the redshifts derived from combining the MCMF outputs of the DES and the high-z runs, $z_\mathrm{comb}$, with those published in \citet{Bocquet19}.
As can be seen, there is good overall agreement for the 
majority of the 481 systems, but there are some outliers.
The scatter between spectroscopic redshifts and MCMF redshifts is consistent with that found in our previous work using ROSAT selected clusters \citep{Klein19}. There are four prominent ($\Delta z > 5 \sigma$) outlier clusters. 
In these four cases, we find two counterparts along the line of sight, where the second ranked one is consistent with the previously published redshift. In all four cases the primary counterpart redshifts are coming from the DES-based run but are consistent with the counterpart from WISE-based MCMF run, making it unlikely that we are observing a 
new failure mode in one of the MCMF runs. One possible explanation for these outliers could be that the original SPT-SZ cluster by cluster follow-up  may be composed of shallow observations that are sufficient to reliably detect the lower redshift counterpart but miss the higher redshift, more significant counterpart. There is further indication that there might be a mild under estimation of the redshifts given in \citet{Bocquet19} for $z>0.7$.

We conclude from the comparison to the previous version of the SPT-SZ cluster that there is consistency for $\sim 99$\% of the overlapping sample. The number of previous systems not making our selection threshold is consistent with our estimate of incompleteness introduced by the optical cleaning, and the most prominent outliers in terms of redshift can be explained as multiple optical systems along the line of sight, where the current analysis finds a more significant richness peak than that selected in the original SPT-SZ follow-up.

 \begin{figure}
\includegraphics[keepaspectratio=true,width=0.97\columnwidth]{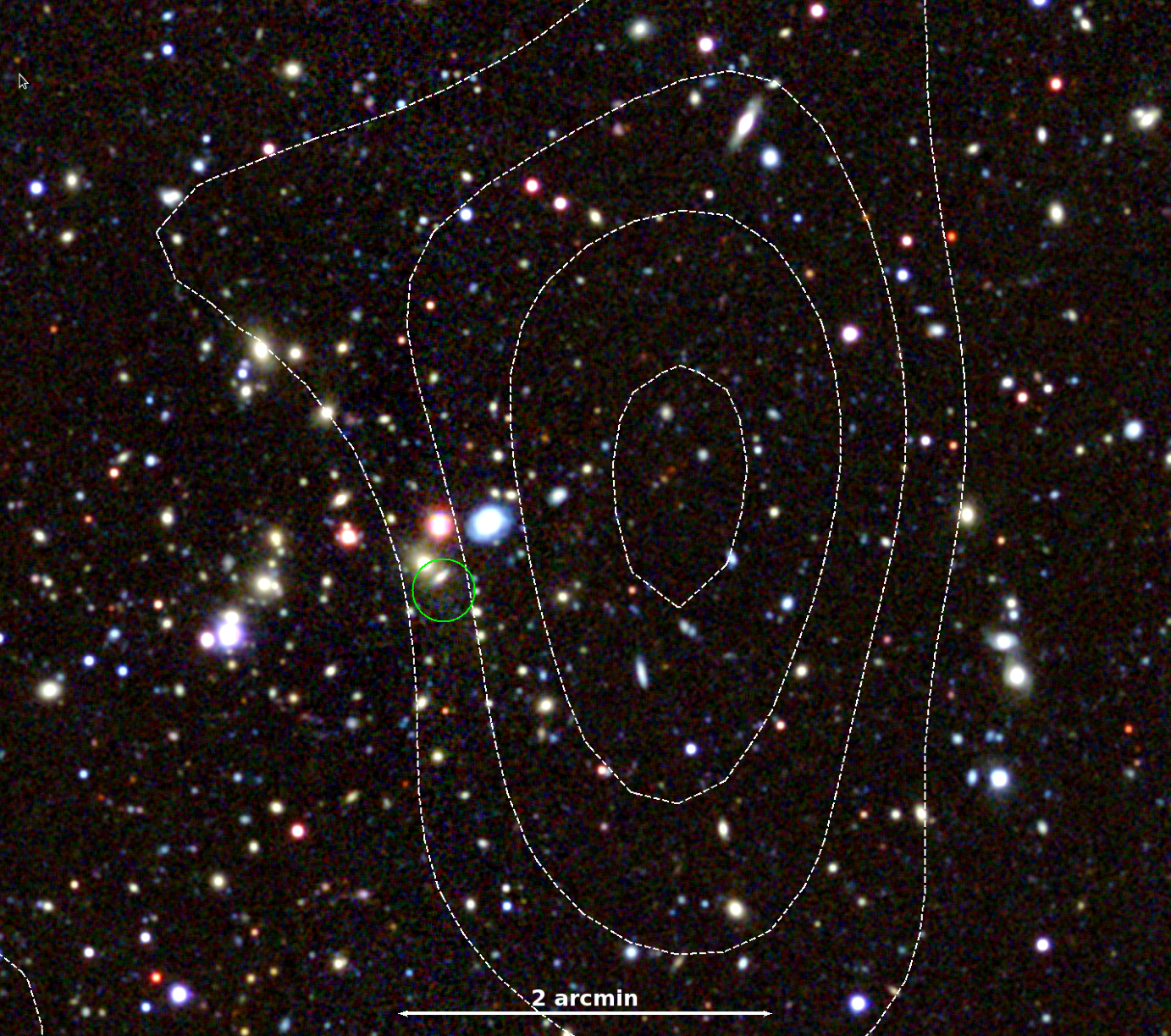}
\includegraphics[keepaspectratio=true,width=0.97\columnwidth]{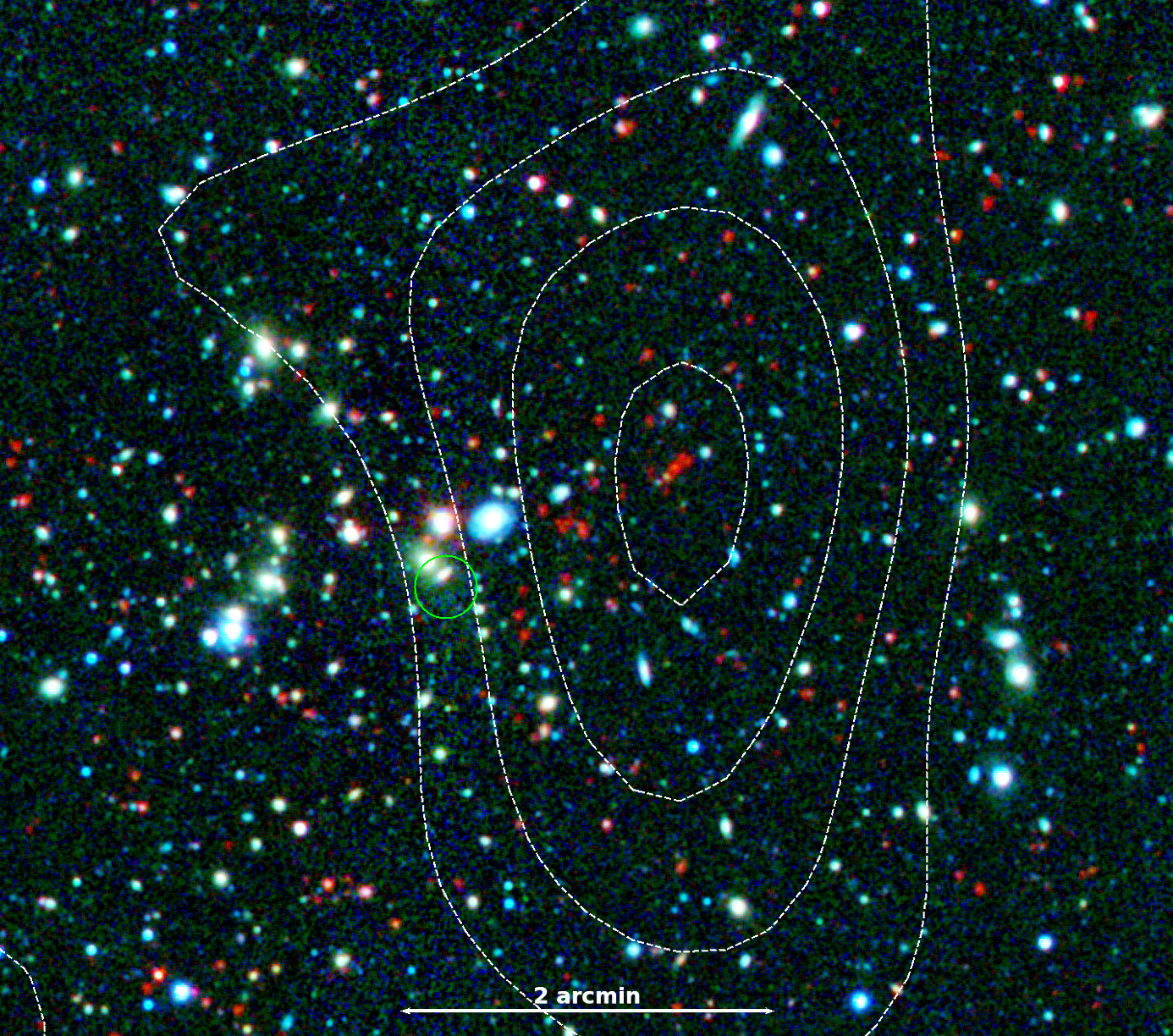}
\vskip-0.10in
\caption{SPT-CL J2331-5736, the cluster with the highest redshift in SPTpol 100d: The top image shows DES $g,r,z$ color composite image, and below is the DES $g,r$ and Spitzer $ch1$ color composite image. The Spitzer image is taken from the SSDF (SPT Spitzer Deep Field). White contours show SPT-SZ S/N contours starting at 1 and increasing in steps of one. The green circle shows the location of a bright radio source detected in SUMSS. MCMF finds two counterparts, the high-z source visible only in the bottom image (\fcont =0.16) and the low-z cluster close to the radio source (\fcont =0.005).}
\label{fig:SPT2331-5736}
\end{figure}

\subsubsection{SPTpol 100d catalog}
The comparison to the SPTpol 100d catalog \citep{SPT100d} is especially interesting, because 
the deeper SPTpol data enable one to identify a larger number of purely SZE-selected clusters, which can then be compared to the MCMF defined catalog from the fully overlapping but shallower SPT-SZ survey data.
This 100d candidate list consists of 89 candidate clusters with a detection S/N $\xi>4.6$. The analysis of image simulations suggests that $81\pm2$ of the candidates are real clusters, which is consistent with the number of optical-IR confirmed systems.

Using a matching radius of 150~arcsec we find 37 matches between the 100d and the SPT-SZ candidate catalogs, with the largest separation being 81~arcsec. Given the fact that contamination of the catalogs is mostly noise driven and the density of contaminants is estimated to be $\approx0.08$\,$\mathrm{deg}^{-2}$ for SPTpol and 0.3--0.4 for SPT-SZ, it is highly unlikely that we would find a chance match within this 150~arcsec search radius.
Therefore, it is safe to assume that all matches correspond to real clusters.

Out of the 37 matches we find 36 with \fcont $< 0.2$ that are members of the SPT-SZ MCMF cluster catalog. The only cluster above that threshold is SPT-CLJ0002--5557, which was discussed in the previous section.  Moreover, missing one cluster out of 37 matches is consistent with the expectation of 2\% incompleteness induced by the optical cleaning undertaken in building the SPT-SZ MCMF catalog. Additionally, we find three clusters with \fcont$ < 0.2$ that were not previously confirmed \citep{SPT100d} and one cluster with a disagreement in redshift.
We discuss those four systems below.

SPT-CL~J2331--5736 (Fig.~\ref{fig:SPT2331-5736}) has a S/N $\xi=4.25$ in SPT-SZ and 8.4 in SPTpol 100d  and is the cluster with the highest redshift in the SPTpol sample with $z=1.38\pm0.1$. In \citet{SPT100d} it is noted that there is also a foreground cluster at $z=0.29$. MCMF finds the low-z cluster to be at $z=0.2975$ with a \fcont$=0.005$ and the high-z structure at $z=1.41$ and \fcont$=0.16$. Given the \fcont\ values, both richness peaks are considered reasonable counterparts in the SPT-SZ MCMF cluster sample.
The low value of \fcont\ makes it highly unlikely that  the low-z structure is a chance superposition near a high-z cluster. The richness of $\lambda=62$ is consistent with the expectation from the scaling relation. On the other hand, the tentative BCG of the high-z cluster is very close to the peak of the SZE signal. A closer investigation reveals a bright radio source with a SUMSS flux of 147.6 mJy (peak, 179.6 total) at the cluster centre of the low-z cluster, which could cause the SZE signal of this cluster to be partially diluted and its centre to be shifted.

\begin{figure}
\includegraphics[keepaspectratio=true,width=0.97\columnwidth]{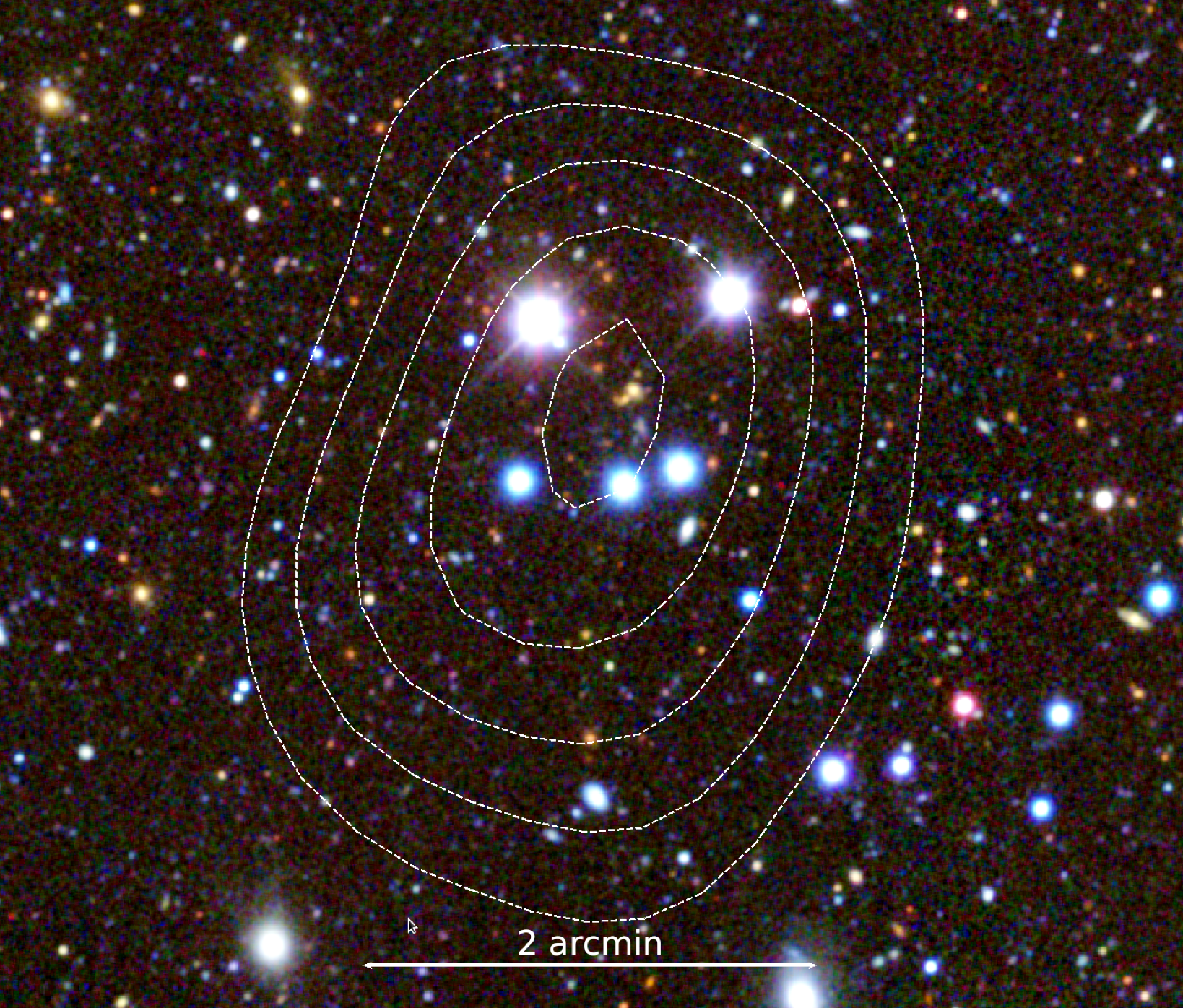}
\vskip-0.10in
\caption{DES $g,r,z$ color composite image of SPT-CL~J2321-5419. White contours show SPT-SZ S/N levels starting at 1 and increasing in steps of one. A bright star makes it difficult to identify the $z=0.79$ cluster members around the star north of the SZE peak.}
\label{fig:SPT2321-5419}
\end{figure}

SPT-CL~J2321--5419 (Fig.~\ref{fig:SPT2321-5419}) has a S/N  $\xi=5.26$ in SPT-SZ and 4.68 in SPTpol 100d . This cluster was not confirmed in the previous SPTpol and SPT-SZ catalogs, because of a bright star close to the SZE postion. The MCMF analysis for this system indicates a redshift of 0.79 and a \fcont$=0.07$. The high-z code finds a consistent redshift but does not confirm this system, due to masking caused by the bright star.
 
 \begin{figure}
\includegraphics[keepaspectratio=true,width=0.97\columnwidth]{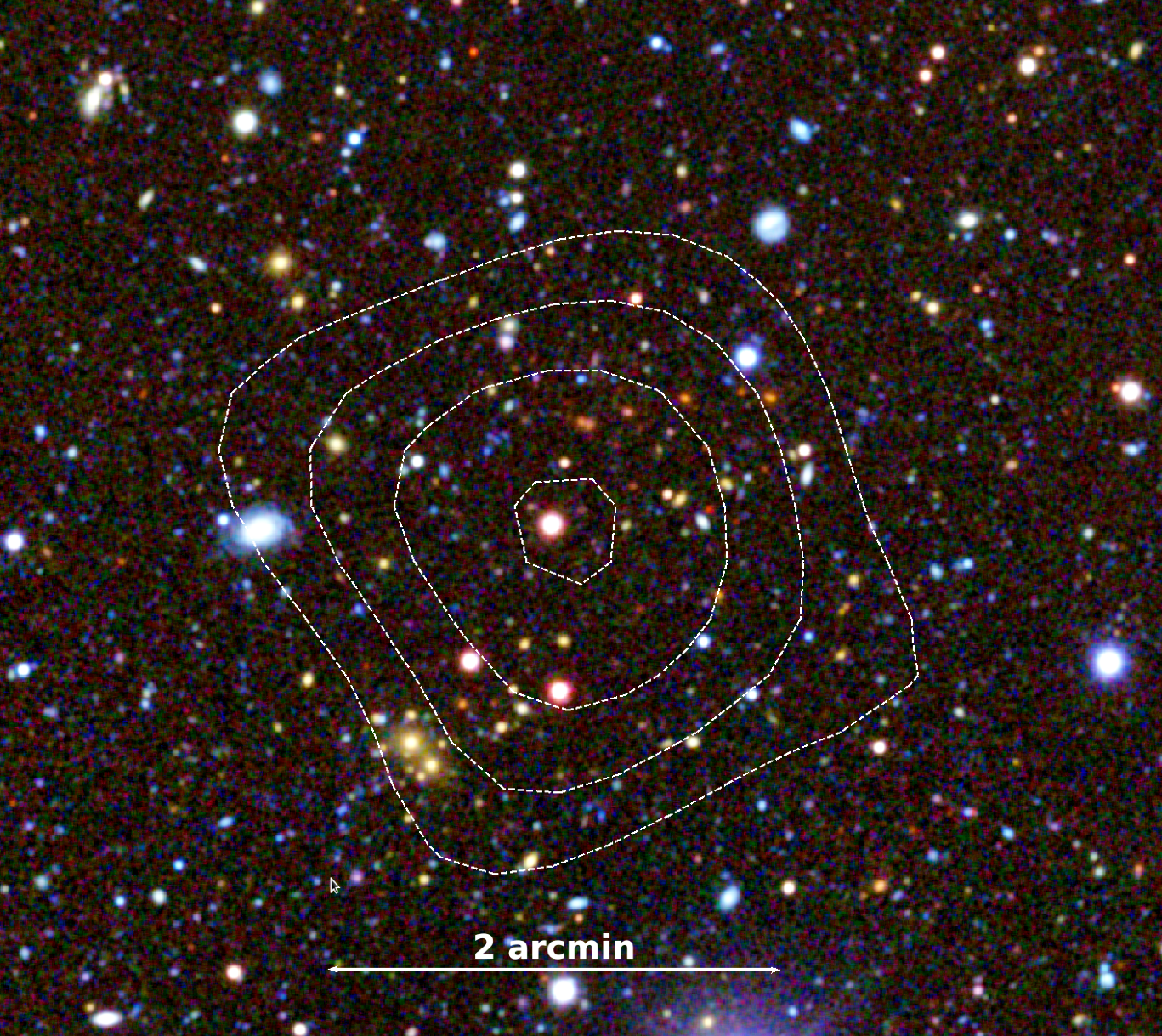}
\vskip-0.10in
\caption{DES $g,r,z$ color composite image of SPT-SZ-CL~J2357-5953, an SPT-SZ to SPTpol 100d match that was not confirmed in SPTpol 100d. White contours show SPT-SZ S/N levels starting at 1 and increasing in steps of one. There are two structures, one at $z=0.517$ and another at $z=1.11$ with corresponding \fcont\ values of 0.02 and 0.18, that are visible to the southeast and northwest of the SZE peak.}
\label{fig:SPT2357-5953}
\end{figure}

SPT-CL~J2357-5953 (Fig.~\ref{fig:SPT2357-5953}) with S/N $\xi=4.13$ in SPT-SZ and 4.66 in SPTpol is unconfirmed in SPTpol 100d, but the MCMF analysis identifies a cluster with redshift $z=0.517$ and \fcont$=0.02$. Additionally, the MCMF analysis identifies a second structure at $z=1.11$ with \fcont$=0.27$.
The peak of the SZE signal is approximately in the middle of the two optical structures, which are separated from each other by 100~arcsec. 
The relatively large separation between the SZE and optical structure positions may have contributed to this system not being confirmed until now. The low probability of having two noise fluctuations in the two SZE surveys agree to within 29~arcsec makes it quite clear that the SZE detection itself is real. The large offset between optical and the SZE centre could be either a result of the low S/N of the detection, or it could be caused by the combination of the SZE signal from both clusters.

 \begin{figure}
\includegraphics[keepaspectratio=true,width=0.97\columnwidth]{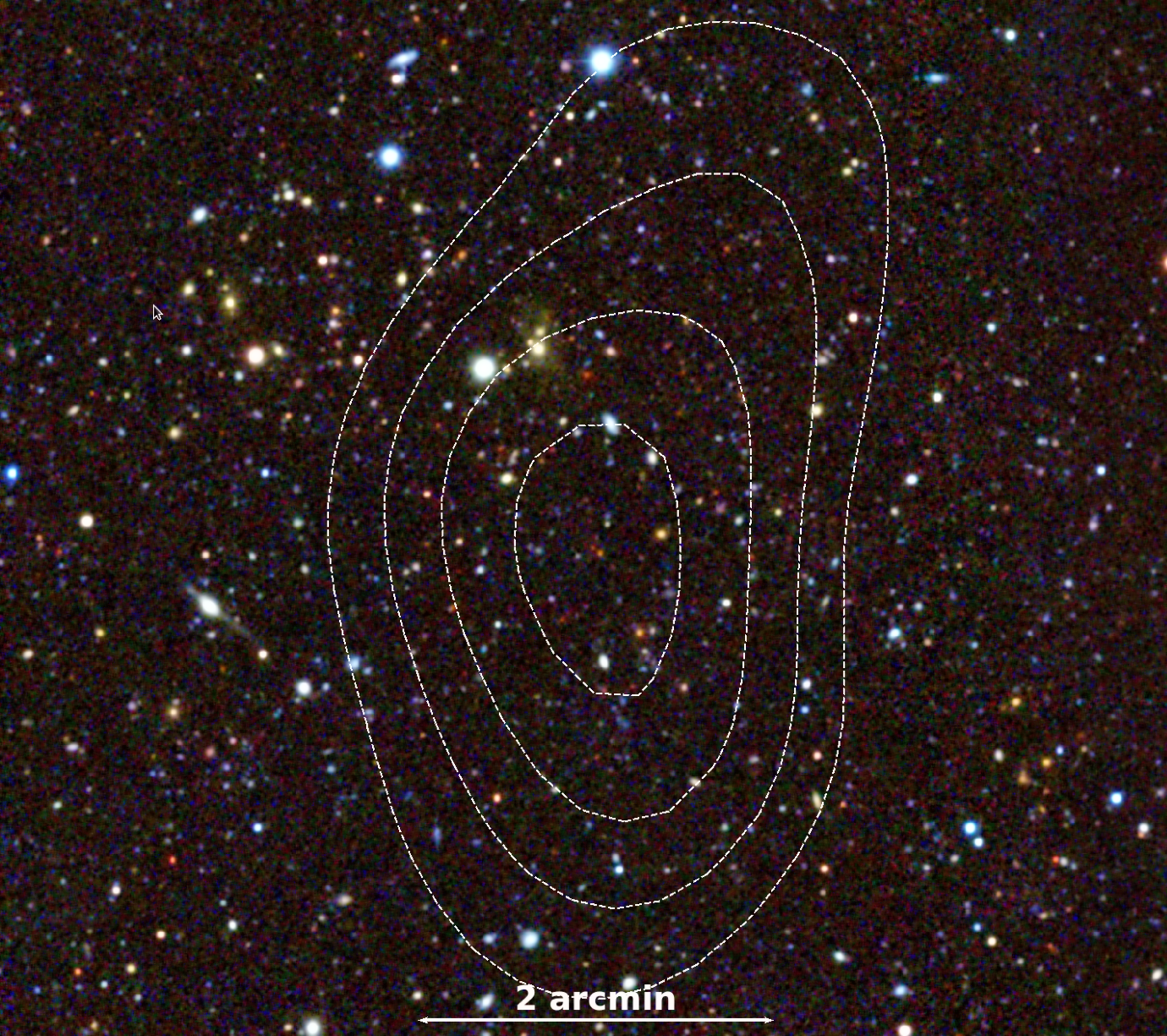}
\includegraphics[keepaspectratio=true,width=0.97\columnwidth]{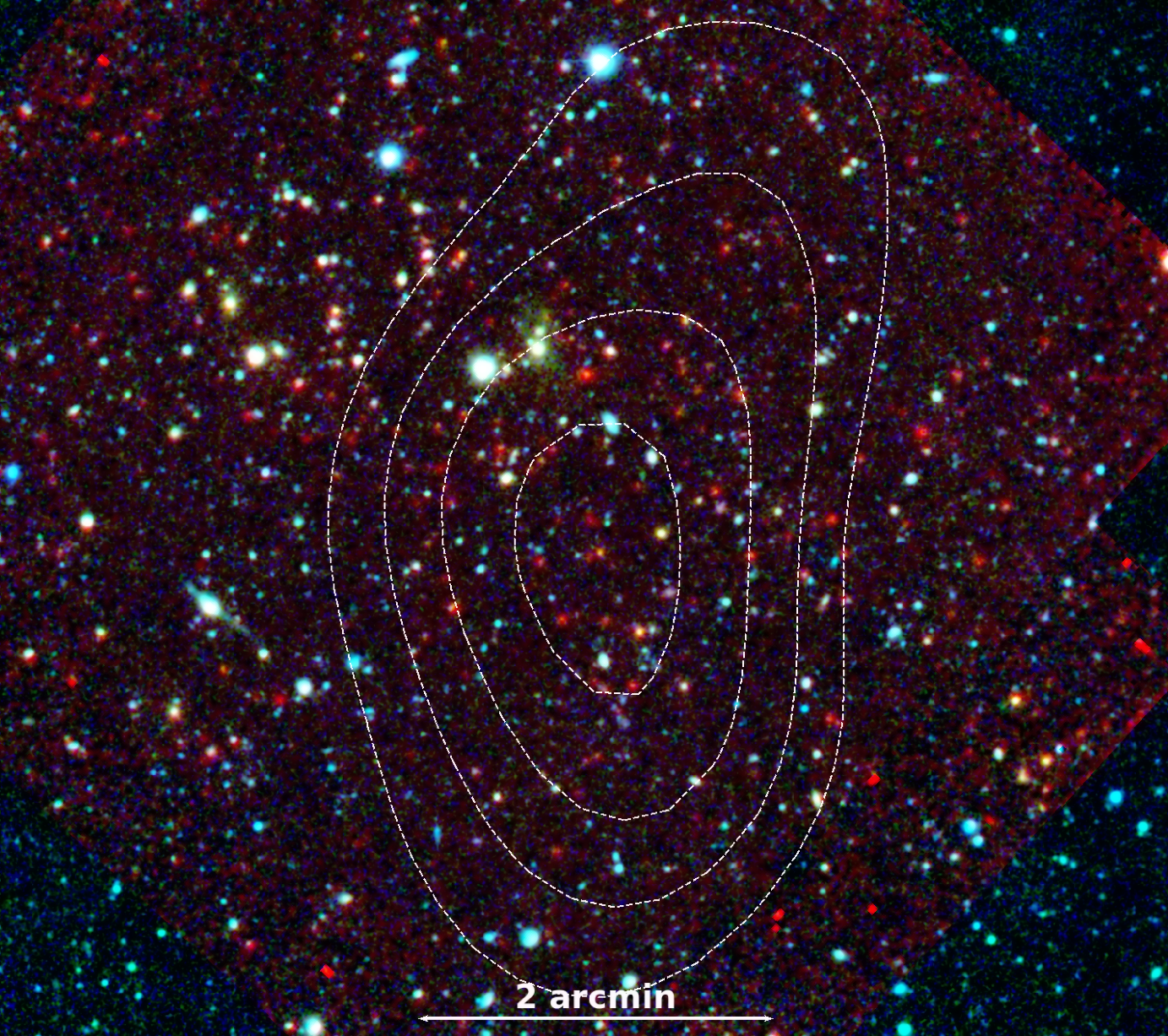}
\vskip-0.10in
\caption{SPT-CL~J0002-5214, an SPT-SZ match to SPTpol 100d not confirmed in SPTpol 100d: Top image shows DES $g,r,z$ color composite image and the bottom image shows the DES $g,r$ and Spitzer $ch1$ color composite image. White contours show SPT-SZ S/N contours starting at 1 and increasing in steps of one. There are two counterparts. One is at $z=0.41$ with \fcont$=0.183$ and another is at $z=1.1$ with \fcont$=0.009$.}
\label{fig:SPT0002-5214}
\end{figure}

The last cluster is SPT-CL~J0002--5214 with S/N $\xi=4.48$ in SPT-SZ and 5.88 in SPTpol. This cluster is listed as a non detection in SPTpol, but there is a note that there is a potential group at $z=0.44$.  Noteworthy here is that according to simulations there should not be any noise fluctuations this large in the SPTpol 100d sample. The analysis with MCMF identifies two structures: one at $z=0.41$ with \fcont$=0.183$ and a second one at $z=1.09$ with \fcont$=0.198$. The high-redshift structure is also independently confirmed by the high-z code with a redshift of $z=1.1$ and \fcont$=0.009$.
Visual inspection of Fig.~\ref{fig:SPT0002-5214} shows the rather compact group at intermediate redshift ($z\sim0.4$), but the high-redshift structure is hard to identify by eye. In the DES $g,r,z$ color composite image there is no clear cluster core, but there are a large number of high-redshift passive galaxies scattered over a region of 1.6~Mpc diameter.
This becomes even clearer when using a combination of DES and Spitzer imaging data.
We therefore conclude that this system is likely a high-redshift cluster with a low optical concentration.

 \begin{figure}
\includegraphics[keepaspectratio=true,width=0.97\columnwidth]{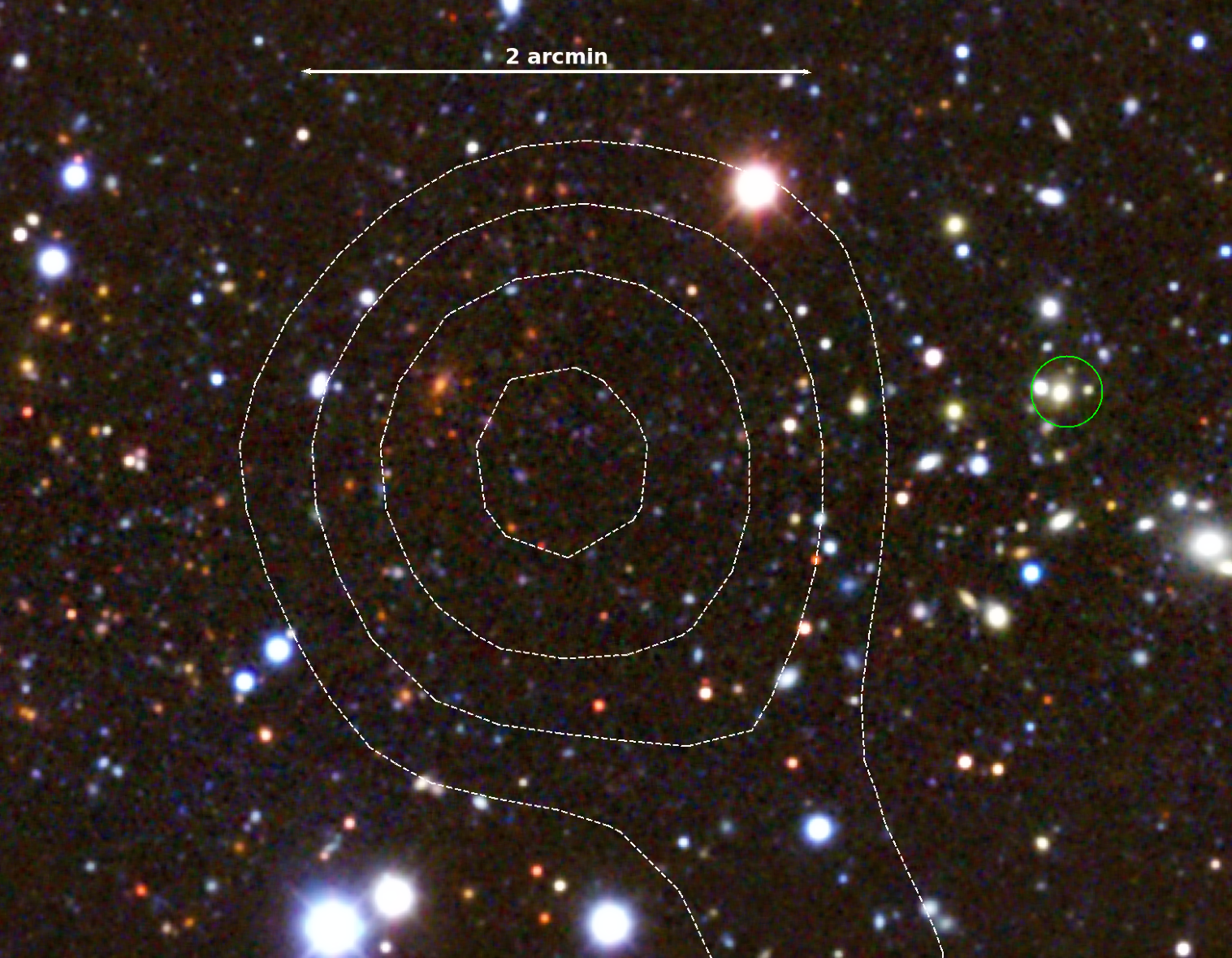}
\vskip-0.10in
\caption{DES $g,r,z$ color composite image of SPT-SZ-CL~J2342-5715. With \fcont$=0.07$ and redshift $z=0.83$ it is the only \fcont$<0.2$ source that does not have a match in SPTpol 100d within the overlapping footprint. White contours show SPT-SZ S/N levels starting at 1 and increasing in steps of one.  The green circle shows the location of a bright radio source detected in SUMSS.}
\label{fig:SPT2342-5715}
\end{figure}
 
 In addition to checking for matched sources as above, we also check for SPT-SZ sources with 
 low \fcont\ that do not appear in the SPTpol 100d catalog.
Because SPTpol 100d  is substantially deeper, we do not expect many SPT-SZ confirmed clusters to be missed, but scatter in both S/N estimates and applied selection thresholds do allow for some number of missed systems. In fact we find just one cluster in the overlapping footprints below \fcont$=0.2$ that is not matched to a SPTpol 100d  source. This source, SPT-SZ-CL~J2342--5715 has a S/N $\xi=4.33$ with \fcont$=0.07$ and a redshift $z=0.83$ (see Fig.~\ref{fig:SPT2342-5715}). The DES optical image reveals a BCG that is only 33~arcsec away from the SZE peak, but the richness of the optical system $\lambda=20.9$ is relatively low.
Within a distance of 1.9~arcminutes we identify a low-z foreground structure harbouring a SUMMS source 
with a flux of $\sim60$ mJy.
Given the \fcont\ value, we can expect to have one contaminating source in the overlapping footprint. At the same time given the scatter in S/N in both surveys, the adopted thresholds in S/N and the low S/N of the particular system one could well find some clusters  at $\xi>4$ in SPT-SZ that are not detected in SPTpol 100d. 
To summarise, we find only one SPT-SZ confirmed system that does not appear in the SPTpol 100d catalog, and given the \fcont\ value this system could indeed be a chance superposition of an SPT-SZ noise fluctuation and an unassociated optical system.

\subsubsection{ACT-DR5 cluster catalog}
The ACT-DR5 cluster catalog \citep{ACTDR5} is an SZE-selected cluster catalog built using ACT survey data.  ACT has similar properties to SPT.  The catalog contains 1,843 clusters over the full DES footprint with  ACT S/N$\ge4$. Allowing for offsets of up to 150~arcsec, we find 415 matches with our SPT-SZ MCMF catalog, where the largest separation is 98~arcsec. Of those matches, 62 clusters have SPT-SZ S/N $\xi<4.5$, and all of them show \fcont$<0.1$, which indicates that these are very likely real clusters. Out of the full overlapping sample of 415, we find two clusters
with \fcont$>0.3$ and one with  $0.2<$\fcont$<0.3$, all of them are known SPT-SZ clusters with $\xi>5$ and would have been considered as confirmed, given the \fcont\ settings tuned for the $\xi>5$ sample. We also find three systems with different redshift estimates.  Two of them indicate two similarly good optical counterparts in the MCMF based analysis where it is the MCMF second ranked system that agrees with the ACT-DR5 redshift. The remaining cluster SPT-CL~J0619--5802, has only one clear MCMF counterpart at $z=0.523$, in agreement with previous SPT-SZ work. The corresponding ACT cluster ACT-CL~J0619.7-5802 is listed with a DES redMaPPer-based redshift of $z=0.391$. Visual inspection supports the MCMF analysis with redshift $z=0.523$.

\begin{figure}
 \includegraphics[keepaspectratio=true,width=1.0\columnwidth]{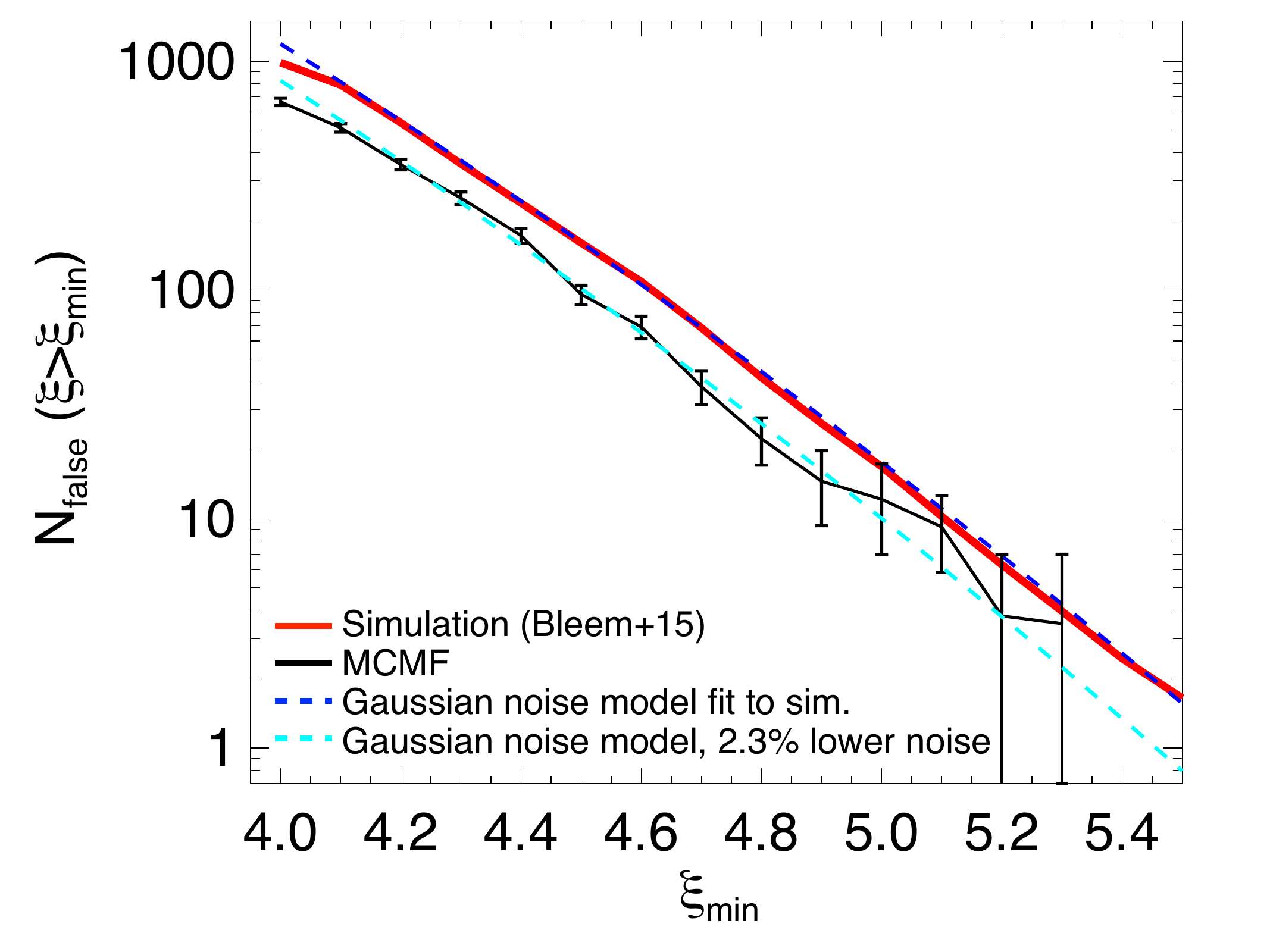}
 \vskip-0.1in
\caption{Cumulative number of contaminants in the SPT-SZ candidate catalog as a function of the SZE signal to noise threshold $\xi_\mathrm{min}$ extracted from image simulations (red) and measured from the SPT-SZ catalog using MCMF-mased mixture model (black line with uncertainties). The simulation-based as well as the MCMF-based estimates can be well described by a Gaussian noise models (dashed blue and cyan lines). The higher number of contaminants in the simulations can be explained by a 2.3\% overestimate of the Gaussian noise used in the image simulations .
See discussion in Section~\ref{sec:contamination}.
}
\label{fig:nfalse}
\end{figure}

 \begin{figure*}
\includegraphics[keepaspectratio=true,width=0.52\linewidth]{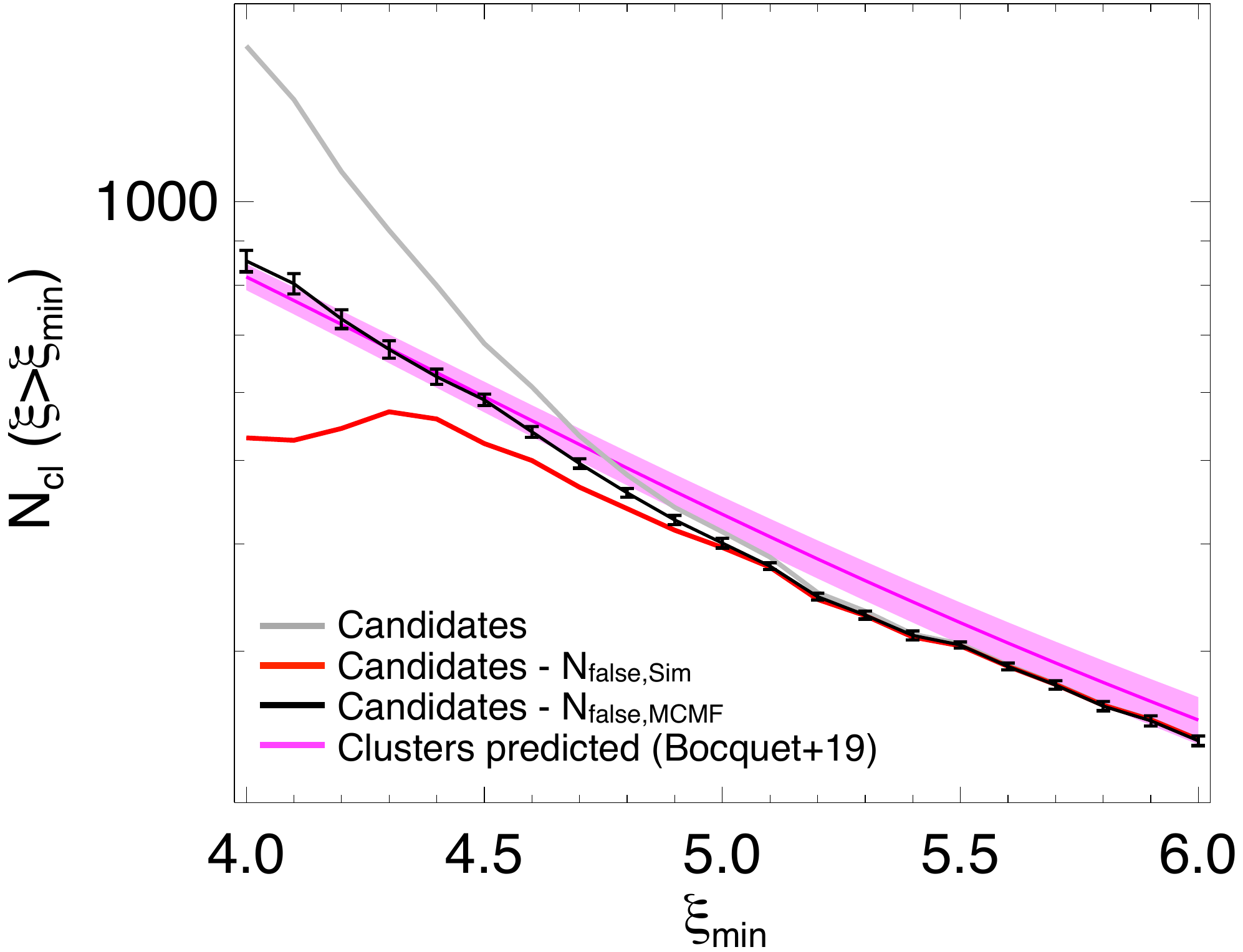}
\includegraphics[keepaspectratio=true,width=0.47\linewidth]{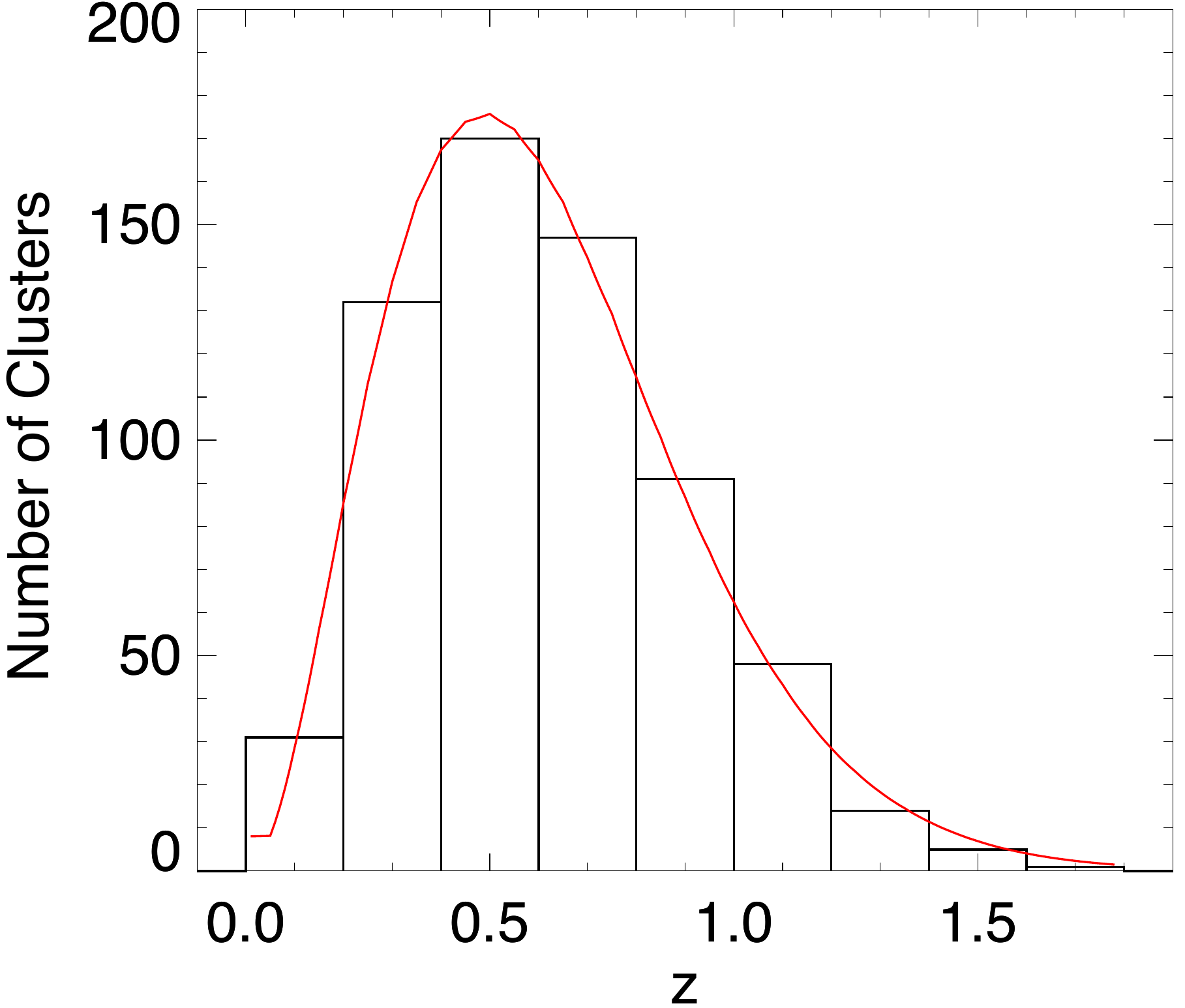}
\vskip-0.10in
\caption{Left: Observed and predicted cluster counts above a given SZE selection threshold $\xi_\mathrm{min}$. All candidates are shown in gray, candidates minus predicted contamination from image simulations in red, clusters expected from the mixture model method (see Figs.~\ref{fig:contaest2} \& ~\ref{fig:puritycomp1}) in black. The predicted number of clusters according to \citet{Bocquet19} is shown in magenta (with 68\% confidence region only includes Poisson noise). Right: Redshift distribution of the $\xi>4.25$ subsample from Table~\ref{tab:sample} (black) and predicted redshift distribution according to \citet{Bocquet19} (red). The predicted counts in $\xi$ space are consistent with the observations, indicating that the sample is an extension of the previous $\xi>5$ sample. The agreement of the shape of the redshift distribution with the prediction suggests that the incompleteness introduced by optical cleaning is not particularly pronounced at any redshift.}
\label{fig:counts}
\end{figure*}

\subsection{Distribution of contaminants in SPT-SZ candidate list}
\label{sec:contaminants}
We can use the MCMF algorithm to estimate the number of contaminants as a function of $\xi$ in the initial SPT-SZ candidate list. Because the SZE is a distinct, negative signal in the 90 and 150~GHz SPT-SZ bandpasses, SZE-selected candidate catalogs contain contamination due to noise fluctuations. Because the noise is close to Gaussian, the number of false detections can be expected to follow a Gaussian noise field. 
The number of contaminants for the SPT-SZ catalog were estimated previously by running the SZE-based cluster finder on source-free simulations \citep{Bleem15}.  The cumulative number of contaminants as a function of S/N $\xi$ is shown in Fig.~\ref{fig:nfalse} together with the best fit model for Gaussian noise (red line and blue dashed line, respectively).  The Gaussian model describes the number of contaminants for $\xi$>4 with two free parameters: 1) the standard deviation of the noise and 2) a normalisation parameter that is related to the ratio of the total survey solid angle to the effective solid angle of the filter functions used to detect clusters in the maps.
This Gaussian model provides an excellent fit to the simulation results.

In the same figure we show the measured number of contaminants extracted using the MCMF-based contamination analysis described in Section~\ref{sec:contamination}.  Interestingly, the shape follows closely that expected from the Gaussian noise model and the image simulations, but the normalization is lower.  In comparison to the Gaussian model fit to the image simulation results, the MCMF-based estimate can be better matched if the standard deviation of the noise is reduced by 2.3\%. Thus, a mild overestimation of the noise in the SPT-SZ data could therefore lead to the overestimation of the contaminants apparent in Fig.~\ref{fig:nfalse}.  We note here that this difference becomes insignificant at $\xi$>5, the threshold of the sample used in previous SPT-SZ cosmological studies, but is large compared to the Poisson uncertainties at $\xi\leq4.7$.

In summary, the distribution of contaminants is consistent with Gaussian noise, as expected, and therefore extremely sensitive to the amplitude of that noise.  There is an offset in the number of contaminants predicted by the image simulations and inferred through the MCMF-based analysis that can be explained by a 2.3\% change in the standard deviation of the noise. In the next section we model the cluster counts and find evidence that points to an overestimate of the contamination in the SPT-SZ candidate list from the image simulations.



\subsection{SPT-SZ MCMF validation using cluster counts}
\label{sec:counts}
Given the new SPT-SZ MCMF cluster sample (Table~\ref{tab:sample}) together with constraints on the residual contamination 
(Section~\ref{sec:contamination}) and incompleteness due to optical cleaning (Section~\ref{sec:incompleteness}), we can obtain the cluster number counts as a function of SPT-SZ S/N $\xi$ threshold ($\xi_\mathrm{min}$), and compare them with the prediction using the results from the cosmological analysis of the previous SPT-SZ sample with $\xi_\mathrm{min}=5$ \citep{Bocquet19}.  Here of course we are mainly interested in the behavior of the new SPT-SZ MCMF clusters with S/N $\xi_\mathrm{min}$<5.   

The expected number of clusters from MCMF-based mixture mode method, as well as from subtracting the simulation-based number of expected false detections from the full list of candidates is shown in Fig.~\ref{fig:counts} (left panel) alongside the predicted number of clusters. Note that the uncertainties shown for the predicted cluster counts represent the Poisson noise only and do not include the error budget due to uncertainties on cosmology and scaling relation parameters. The uncertainties for the optical method predominantly depend on the number of contaminants in the sample and therefore becomes small at high $\xi_\mathrm{min}$.
As can be seen in Fig.~\ref{fig:counts}, the predicted number of clusters shown in magenta agrees with the observed number using the optical method at the $1 \sigma$ level and the behaviour at  S/N $\xi_\mathrm{min}$<5 appears to be a meaningful extension to the high $\xi_\mathrm{min}$ regime. By contrast, the number of clusters expected from using the simulation-based appears to decrease at lower $\xi_\mathrm{min}$, supporting the picture of a mild overestimation of the noise level in the simulations. The difference between simulation-based and optical-based estimates becomes insignificant at $\xi_\mathrm{min}= 5$, which was used in previous SPT-SZ-based cosmological studies.

Using the results from \citet{Bocquet19}, we can further compare the observed and the expected redshift distributions, which we present in the right panel of Fig.~\ref{fig:counts}. Here we use the $\xi>4.25$ subsample, which--- according to 
Fig.~\ref{fig:complete}--- is 96\% pure and 96.5\% complete with respect to the initial SZE candidate selection. By design the \fcont-based selection aims to maintain a constant level of contamination as a function of redshift. Contamination therefore should not alter the shape of the redshift distribution of the sample. The red line in the right panel of Fig.~\ref{fig:counts} shows the predicted redshift distribution using the results from \citet{Bocquet19} normalised to same total number of clusters. The predicted and observed shapes of the redshift distributions agree remarkably well. Under the assumption that the contamination fraction is indeed constant over redshift this suggests that the incompleteness introduced by the \fcont$ <0.2$ selection is not significantly impacting the redshift distribution either.


\section{Conclusions}
\label{sec:conclusions}
In this paper, we present the SPT-SZ MCMF cluster catalog with candidates selected to have SPT-SZ S/N $\xi>4$ that are then confirmed using the MCMF algorithm. This sample represents a $\approx$ 50\% increase in size
compared to the previous SPT-SZ catalog and contains 811 clusters with 9\% contamination. Subsamples of this new catalog can be selected to have different characteristics (see Table~\ref{tab:sample}).  Considering an SPT-SZ S/N threshold $\xi>4.25$ with stricter \fcont\ constraints in order to remove chance superpositions (\fcont$<0.125$), we obtain 640 clusters with 96\% purity.  This subsample has a modest 3.5\% incompleteness due to optical cleaning with the MCMF algorithm. 
This sample should meet the requirements for a cosmological analysis and corresponds to a factor two increase compared to the previous SPT-SZ cluster catalog used for cosmological analysis \citep{Bocquet19}. 

We use information derived from our MCMF-based analysis to infer the level of the initial contamination in the SZE-selected sample above several S/N thresholds as well as the purity and completeness after optical confirmation. This information can be used to select the combination of purity, sample size and completeness best suited for a given science study. Studies less impacted by contamination or that suffer from small number statistics may chose larger but more contaminated subsamples, while studies sensitive on contamination may use cleaner but smaller subsamples. The measured initial contamination, expressed in number of false detections above a S/N threshold, follows the shape expected for Gaussian noise. We find a systematic difference between our measurements and those predicted by simulations that could be explained if the noise assumed in the simulation was overestimated by a small amount (2.3\%). Comparing number of false detections with number of candidates we find further evidence that the simulation-based estimates over predict the number of false detections as the number of real systems ($N_\mathrm{cand}-N_\mathrm{false}$) above a S/N threshold appears to decrease when lowering the threshold.

A validation test consisting of the comparison of S/N $\xi$ and redshift  $z$ distributions of the new SPT-SZ MCMF sample to the predictions extrapolated from the previous cosmological analysis of the $\xi>5$ subsample \citep{Bocquet19} shows good agreement.  This gives us confidence that the new sample is well suited for an updated cosmological analysis that will be carried out in combination with the DES weak lensing dataset to constrain cluster masses (Bocquet et al, in prep). The subsample anticipated for that study is the more conservative subsample that contains 480 clusters with SPT-SZ S/N $\xi>4.5$ and $z>0.25$ (see Table~\ref{tab:sample}).

Combining the SPT-SZ sample with SPT-ECS \citep{Bleem20} and SPTpol 100d \citep{SPT100d} the total number of confirmed SPT-selected clusters now raises to 1,343. This number will further rise with the soon to be published sample from SPTpol 500d (Bleem et al., in prep).



\section*{Acknowledgements}
We acknowledge the support of the Max Planck Society Faculty Fellowship program at MPE, the support of the DFG (German Research Foundation) Cluster of Excellence ``Origin and Structure of the 
Universe'' - EXC-2094 - 390783311, and the Ludwig-Maximilians-Universit\"at in Munich.

Funding for the DES Projects has been provided by the U.S. Department of Energy, the U.S. National Science 
Foundation, the Ministry of Science and Education of Spain, the Science and Technology Facilities Council of the 
United Kingdom, the Higher Education Funding Council for England, the National Center for Supercomputing  
Applications at the University of Illinois at Urbana-Champaign, the Kavli Institute of Cosmological Physics at the 
University of Chicago, the Center for Cosmology and Astro-Particle Physics at the Ohio State University, the 
Mitchell Institute for Fundamental Physics and Astronomy at Texas A\&M University, Financiadora de Estudos e 
Projetos, Funda{\c c}{\~a}o Carlos Chagas Filho de Amparo {\`a} Pesquisa do Estado do Rio de Janeiro, Conselho 
Nacional de Desenvolvimento Cient{\'i}fico e Tecnol{\'o}gico and the Minist{\'e}rio da Ci{\^e}ncia, Tecnologia e 
Inova{\c c}{\~a}o, the Deutsche Forschungsgemeinschaft and the Collaborating Institutions in the Dark Energy 
Survey.   The Collaborating Institutions are Argonne National Laboratory, the University of California at Santa Cruz, 
the University of Cambridge, Centro de Investigaciones Energ{\'e}ticas, Medioambientales y Tecnol{\'o}gicas-
Madrid, the University of Chicago, University College London, the DES-Brazil Consortium, the University of 
Edinburgh, the Eidgen{\"o}ssische Technische Hochschule (ETH) Z{\"u}rich, Fermi National Accelerator Laboratory, 
the University of Illinois at Urbana-Champaign, the Institut de Ci{\`e}ncies de l'Espai (IEEC/CSIC), the Institut de 
F{\'i}sica d'Altes Energies, Lawrence Berkeley National Laboratory, the Ludwig-Maximilians Universit{\"a}t M{\"u}nchen and the associated Excellence Cluster Origins, the University of Michigan, the National Optical Astronomy 
Observatory, the University of Nottingham, The Ohio State University, the University of Pennsylvania, the 
University of Portsmouth, SLAC National Accelerator Laboratory, Stanford University, the University of Sussex, 
Texas A\&M University, and the OzDES Membership Consortium.  The DES data management system is 
supported by the National Science Foundation under Grant Number AST-1138766. The DES participants from 
Spanish institutions are partially supported by MINECO under grants AYA2012-39559, ESP2013-48274, 
FPA2013-47986, and Centro de Excelencia Severo Ochoa SEV-2012-0234.  Research leading to these results has 
received funding from the European Research Council under the European Union's Seventh Framework 
Programme (FP7/2007-2013) including ERC grant agreements  240672, 291329, and 306478.

The South Pole Telescope program is supported by the National Science Foundation (NSF) through the Grant No. OPP-1852617. Partial support is also provided by the Kavli Institute of Cosmological Physics at the University of Chicago.  Argonne National Laboratory’s work was supported by the U.S. Department of Energy, Office of High Energy Physics, under Contract No. DE-AC02-06CH11357. Work at Fermi National Accelerator Laboratory, a DOE-OS, HEP User Facility managed by the Fermi Research Alliance, LLC, was supported under Contract No. DE-AC02-07CH11359.

MC, LS, AS, and VS are supported by the ERC-StG ‘ClustersXCosmo’ grant agreement 716762. AS is further supported by the FARE-MIUR grant 'ClustersXEuclid' R165SBKTMA, by INFN InDark Grant, and by the ICSC National Recovery and Resilience Plan (PNRR) Project ID CN00000013 "Italian Research Center on High-Performance Computing, Big Data and Quantum Computing" funded by MUR Missione 4 Componente 2 Investimento 1.4 - Next Generation EU (NGEU).



\bibliographystyle{mnras}
\bibliography{optid_refs}

\begin{thebibliography}{}
\makeatletter
\relax
\def\mn@urlcharsother{\let\do\@makeother \do\$\do\&\do\#\do\^\do\_\do\%\do\~}
\def\mn@doi{\begingroup\mn@urlcharsother \@ifnextchar [ {\mn@doi@}
  {\mn@doi@[]}}
\def\mn@doi@[#1]#2{\def\@tempa{#1}\ifx\@tempa\@empty \href
  {http://dx.doi.org/#2} {doi:#2}\else \href {http://dx.doi.org/#2} {#1}\fi
  \endgroup}
\def\mn@eprint#1#2{\mn@eprint@#1:#2::\@nil}
\def\mn@eprint@arXiv#1{\href {http://arxiv.org/abs/#1} {{\tt arXiv:#1}}}
\def\mn@eprint@dblp#1{\href {http://dblp.uni-trier.de/rec/bibtex/#1.xml}
  {dblp:#1}}
\def\mn@eprint@#1:#2:#3:#4\@nil{\def\@tempa {#1}\def\@tempb {#2}\def\@tempc
  {#3}\ifx \@tempc \@empty \let \@tempc \@tempb \let \@tempb \@tempa \fi \ifx
  \@tempb \@empty \def\@tempb {arXiv}\fi \@ifundefined
  {mn@eprint@\@tempb}{\@tempb:\@tempc}{\expandafter \expandafter \csname
  mn@eprint@\@tempb\endcsname \expandafter{\@tempc}}}

\bibitem[\protect\citeauthoryear{{Abbott} et~al.,}{{Abbott}
  et~al.}{2018}]{DESDR1}
{Abbott} T.~M.~C.,  et~al., 2018, \mn@doi [\apjs] {10.3847/1538-4365/aae9f0},
  \href {https://ui.adsabs.harvard.edu/abs/2018ApJS..239...18A} {239, 18}

\bibitem[\protect\citeauthoryear{{Aihara} et~al.,}{{Aihara}
  et~al.}{2018}]{HSC-SSP}
{Aihara} H.,  et~al., 2018, \mn@doi [\pasj] {10.1093/pasj/psx066}, \href
  {https://ui.adsabs.harvard.edu/abs/2018PASJ...70S...4A} {70, S4}

\bibitem[\protect\citeauthoryear{{Benson} et~al.,}{{Benson}
  et~al.}{2013}]{Benson13}
{Benson} B.~A.,  et~al., 2013, \mn@doi [\apj] {10.1088/0004-637X/763/2/147},
  \href {https://ui.adsabs.harvard.edu/abs/2013ApJ...763..147B} {763, 147}

\bibitem[\protect\citeauthoryear{{Bertin} \& {Arnouts}}{{Bertin} \&
  {Arnouts}}{1996}]{bertin96}
{Bertin} E.,  {Arnouts} S.,  1996, \aaps, \href
  {http://adsabs.harvard.edu/abs/1996A%26AS..117..393B} {117, 393}

\bibitem[\protect\citeauthoryear{{Blanton} et~al.,}{{Blanton}
  et~al.}{2017}]{SDSS}
{Blanton} M.~R.,  et~al., 2017, \mn@doi [\aj] {10.3847/1538-3881/aa7567}, \href
  {https://ui.adsabs.harvard.edu/abs/2017AJ....154...28B} {154, 28}

\bibitem[\protect\citeauthoryear{{Bleem} et~al.,}{{Bleem}
  et~al.}{2015}]{Bleem15}
{Bleem} L.~E.,  et~al., 2015, \mn@doi [\apjs] {10.1088/0067-0049/216/2/27},
  \href {https://ui.adsabs.harvard.edu/abs/2015ApJS..216...27B} {216, 27}

\bibitem[\protect\citeauthoryear{{Bleem} et~al.,}{{Bleem}
  et~al.}{2020}]{Bleem20}
{Bleem} L.~E.,  et~al., 2020, \mn@doi [\apjs] {10.3847/1538-4365/ab6993}, \href
  {https://ui.adsabs.harvard.edu/abs/2020ApJS..247...25B} {247, 25}

\bibitem[\protect\citeauthoryear{{Bocquet} et~al.,}{{Bocquet}
  et~al.}{2015}]{Bocquet15}
{Bocquet} S.,  et~al., 2015, \mn@doi [\apj] {10.1088/0004-637X/799/2/214},
  \href {https://ui.adsabs.harvard.edu/abs/2015ApJ...799..214B} {799, 214}

\bibitem[\protect\citeauthoryear{{Bocquet} et~al.,}{{Bocquet}
  et~al.}{2019}]{Bocquet19}
{Bocquet} S.,  et~al., 2019, \mn@doi [\apj] {10.3847/1538-4357/ab1f10}, \href
  {https://ui.adsabs.harvard.edu/abs/2019ApJ...878...55B} {878, 55}

\bibitem[\protect\citeauthoryear{{B{\"o}hringer} et~al.,}{{B{\"o}hringer}
  et~al.}{2000}]{NORAS}
{B{\"o}hringer} H.,  et~al., 2000, \mn@doi [\apjs] {10.1086/313427}, \href
  {https://ui.adsabs.harvard.edu/abs/2000ApJS..129..435B} {129, 435}

\bibitem[\protect\citeauthoryear{{Brunner} et~al.,}{{Brunner}
  et~al.}{2022}]{Brunner21}
{Brunner} H.,  et~al., 2022, \mn@doi [\aap] {10.1051/0004-6361/202141266},
  \href {https://ui.adsabs.harvard.edu/abs/2022A&A...661A...1B} {661, A1}

\bibitem[\protect\citeauthoryear{{Carlstrom} et~al.,}{{Carlstrom}
  et~al.}{2011}]{Carlstrom11}
{Carlstrom} J.~E.,  et~al., 2011, \mn@doi [\pasp] {10.1086/659879}, \href
  {https://ui.adsabs.harvard.edu/abs/2011PASP..123..568C} {123, 568}

\bibitem[\protect\citeauthoryear{{Chiu}, {Klein}, {Mohr}  \& {Bocquet}}{{Chiu}
  et~al.}{2023}]{Chiu23}
{Chiu} I.~N.,  {Klein} M.,  {Mohr} J.,   {Bocquet} S.,  2023, \mn@doi [\mnras]
  {10.1093/mnras/stad957}, \href
  {https://ui.adsabs.harvard.edu/abs/2023MNRAS.tmp..949C} {}

\bibitem[\protect\citeauthoryear{{Cutri} et~al.,}{{Cutri}
  et~al.}{2013}]{Cutri13}
{Cutri} R.~M.,  et~al., 2013, {Explanatory Supplement to the AllWISE Data
  Release Products}, Explanatory Supplement to the AllWISE Data Release
  Products, by R. M. Cutri et al.

\bibitem[\protect\citeauthoryear{{Dey} et~al.,}{{Dey}
  et~al.}{2019}]{Legacysurveys19}
{Dey} A.,  et~al., 2019, \mn@doi [\aj] {10.3847/1538-3881/ab089d}, \href
  {https://ui.adsabs.harvard.edu/abs/2019AJ....157..168D} {157, 168}

\bibitem[\protect\citeauthoryear{{Drlica-Wagner} et~al.,}{{Drlica-Wagner}
  et~al.}{2018}]{DESY1Gold}
{Drlica-Wagner} A.,  et~al., 2018, \mn@doi [\apjs] {10.3847/1538-4365/aab4f5},
  \href {http://adsabs.harvard.edu/abs/2018ApJS..235...33D} {235, 33}

\bibitem[\protect\citeauthoryear{{Finoguenov} et~al.,}{{Finoguenov}
  et~al.}{2020}]{CODEX19}
{Finoguenov} A.,  et~al., 2020, \mn@doi [\aap] {10.1051/0004-6361/201937283},
  \href {https://ui.adsabs.harvard.edu/abs/2020A&A...638A.114F} {638, A114}

\bibitem[\protect\citeauthoryear{{Flaugher} et~al.,}{{Flaugher}
  et~al.}{2015}]{Flaugher15}
{Flaugher} B.,  et~al., 2015, \mn@doi [\aj] {10.1088/0004-6256/150/5/150},
  \href {http://adsabs.harvard.edu/abs/2015AJ....150..150F} {150, 150}

\bibitem[\protect\citeauthoryear{{Gladders} \& {Yee}}{{Gladders} \&
  {Yee}}{2000}]{gladders00}
{Gladders} M.~D.,  {Yee} H. K.~C.,  2000, \aj, \href
  {http://adsabs.harvard.edu/cgi-bin/nph-bib_query?bibcode=2000AJ....120.2148G&db_key=AST}
  {120, 2148}

\bibitem[\protect\citeauthoryear{{Gonzalez} et~al.,}{{Gonzalez}
  et~al.}{2019}]{Madcows}
{Gonzalez} A.~H.,  et~al., 2019, \mn@doi [\apjs] {10.3847/1538-4365/aafad2},
  \href {https://ui.adsabs.harvard.edu/abs/2019ApJS..240...33G} {240, 33}

\bibitem[\protect\citeauthoryear{{Grandis} et~al.,}{{Grandis}
  et~al.}{2020}]{Grandis20}
{Grandis} S.,  et~al., 2020, \mn@doi [\mnras] {10.1093/mnras/staa2333}, \href
  {https://ui.adsabs.harvard.edu/abs/2020MNRAS.498..771G} {498, 771}

\bibitem[\protect\citeauthoryear{{Hern{\'a}ndez-Lang}
  et~al.,}{{Hern{\'a}ndez-Lang} et~al.}{2023}]{Hernandez22}
{Hern{\'a}ndez-Lang} D.,  et~al., 2023, \mn@doi [\mnras]
  {10.1093/mnras/stad2319}, \href
  {https://ui.adsabs.harvard.edu/abs/2023MNRAS.525...24H} {525, 24}

\bibitem[\protect\citeauthoryear{{Hilton} et~al.,}{{Hilton}
  et~al.}{2021}]{ACTDR5}
{Hilton} M.,  et~al., 2021, \mn@doi [\apjs] {10.3847/1538-4365/abd023}, \href
  {https://ui.adsabs.harvard.edu/abs/2021ApJS..253....3H} {253, 3}

\bibitem[\protect\citeauthoryear{{Huang} et~al.,}{{Huang}
  et~al.}{2020}]{SPT100d}
{Huang} N.,  et~al., 2020, \mn@doi [\aj] {10.3847/1538-3881/ab6a96}, \href
  {https://ui.adsabs.harvard.edu/abs/2020AJ....159..110H} {159, 110}

\bibitem[\protect\citeauthoryear{{King}}{{King}}{1962}]{king62}
{King} I.,  1962, \mn@doi [\aj] {10.1086/108756}, \href
  {http://cdsads.u-strasbg.fr/abs/1962AJ.....67..471K} {67, 471}

\bibitem[\protect\citeauthoryear{{Klein} et~al.,}{{Klein}
  et~al.}{2018}]{Klein18}
{Klein} M.,  et~al., 2018, \mn@doi [\mnras] {10.1093/mnras/stx2929}, \href
  {https://ui.adsabs.harvard.edu/abs/2018MNRAS.474.3324K} {474, 3324}

\bibitem[\protect\citeauthoryear{{Klein} et~al.,}{{Klein}
  et~al.}{2019}]{Klein19}
{Klein} M.,  et~al., 2019, \mn@doi [\mnras] {10.1093/mnras/stz1463}, \href
  {https://ui.adsabs.harvard.edu/abs/2019MNRAS.488..739K} {488, 739}

\bibitem[\protect\citeauthoryear{{Klein} et~al.,}{{Klein}
  et~al.}{2022}]{Klein22}
{Klein} M.,  et~al., 2022, \mn@doi [\aap] {10.1051/0004-6361/202141123}, \href
  {https://ui.adsabs.harvard.edu/abs/2022A&A...661A...4K} {661, A4}

\bibitem[\protect\citeauthoryear{{Klein}, {Hern{\'a}ndez-Lang}, {Mohr}  \&
  {Singh}}{{Klein} et~al.}{2023}]{Klein23}
{Klein} M.,  {Hern{\'a}ndez-Lang} D.,  {Mohr} J.~J.,   {Singh} A.,  2023,
  \mn@doi [arXiv e-prints] {10.48550/arXiv.2305.20066}, \href
  {https://ui.adsabs.harvard.edu/abs/2023arXiv230520066K} {p. arXiv:2305.20066}

\bibitem[\protect\citeauthoryear{{Lahav}, {Edge}, {Fabian}  \&
  {Putney}}{{Lahav} et~al.}{1989}]{Lahav89}
{Lahav} O.,  {Edge} A.~C.,  {Fabian} A.~C.,   {Putney} A.,  1989, \mn@doi
  [\mnras] {10.1093/mnras/238.3.881}, \href
  {https://ui.adsabs.harvard.edu/abs/1989MNRAS.238..881L} {238, 881}

\bibitem[\protect\citeauthoryear{{Lang}}{{Lang}}{2014}]{unWISE14}
{Lang} D.,  2014, \mn@doi [\aj] {10.1088/0004-6256/147/5/108}, \href
  {https://ui.adsabs.harvard.edu/abs/2014AJ....147..108L} {147, 108}

\bibitem[\protect\citeauthoryear{{Mainzer} et~al.,}{{Mainzer}
  et~al.}{2011}]{Mainzer11}
{Mainzer} A.,  et~al., 2011, \mn@doi [\apj] {10.1088/0004-637X/731/1/53}, \href
  {https://ui.adsabs.harvard.edu/abs/2011ApJ...731...53M} {731, 53}

\bibitem[\protect\citeauthoryear{{Mantz}, {Allen}, {Rapetti}  \&
  {Ebeling}}{{Mantz} et~al.}{2010}]{Mantz10}
{Mantz} A.,  {Allen} S.~W.,  {Rapetti} D.,   {Ebeling} H.,  2010, \mn@doi
  [\mnras] {10.1111/j.1365-2966.2010.16992.x}, \href
  {https://ui.adsabs.harvard.edu/abs/2010MNRAS.406.1759M} {406, 1759}

\bibitem[\protect\citeauthoryear{{Meisner}, {Lang}, {Schlafly}  \&
  {Schlegel}}{{Meisner} et~al.}{2019}]{unWISE19}
{Meisner} A.~M.,  {Lang} D.,  {Schlafly} E.~F.,   {Schlegel} D.~J.,  2019,
  \mn@doi [\pasp] {10.1088/1538-3873/ab3df4}, \href
  {https://ui.adsabs.harvard.edu/abs/2019PASP..131l4504M} {131, 124504}

\bibitem[\protect\citeauthoryear{{Merloni} et~al.,}{{Merloni}
  et~al.}{2012}]{Merloni12}
{Merloni} A.,  et~al., 2012, arXiv e-prints, \href
  {https://ui.adsabs.harvard.edu/abs/2012arXiv1209.3114M} {p. arXiv:1209.3114}

\bibitem[\protect\citeauthoryear{{Mohr}, {Fabricant}  \& {Geller}}{{Mohr}
  et~al.}{1993}]{mohr93}
{Mohr} J.~J.,  {Fabricant} D.~G.,   {Geller} M.~J.,  1993, \mn@doi [\apj]
  {10.1086/173019}, \href
  {https://ui.adsabs.harvard.edu/abs/1993ApJ...413..492M} {413, 492}

\bibitem[\protect\citeauthoryear{{Morganson} et~al.,}{{Morganson}
  et~al.}{2018}]{Morganson18}
{Morganson} E.,  et~al., 2018, \mn@doi [\pasp] {10.1088/1538-3873/aab4ef},
  \href {https://ui.adsabs.harvard.edu/abs/2018PASP..130g4501M} {130, 074501}

\bibitem[\protect\citeauthoryear{{Muzzin} et~al.,}{{Muzzin}
  et~al.}{2009}]{Muzzin09}
{Muzzin} A.,  et~al., 2009, \mn@doi [\apj] {10.1088/0004-637X/698/2/1934},
  \href {https://ui.adsabs.harvard.edu/abs/2009ApJ...698.1934M} {698, 1934}

\bibitem[\protect\citeauthoryear{{Nurgaliev}, {McDonald}, {Benson}, {Miller},
  {Stubbs}  \& {Vikhlinin}}{{Nurgaliev} et~al.}{2013}]{nurgaliev13}
{Nurgaliev} D.,  {McDonald} M.,  {Benson} B.~A.,  {Miller} E.~D.,  {Stubbs}
  C.~W.,   {Vikhlinin} A.,  2013, \mn@doi [\apj] {10.1088/0004-637X/779/2/112},
  \href {http://adsabs.harvard.edu/abs/2013ApJ...779..112N} {779, 112}

\bibitem[\protect\citeauthoryear{{Nurgaliev} et~al.,}{{Nurgaliev}
  et~al.}{2017}]{nurgaliev17}
{Nurgaliev} D.,  et~al., 2017, \mn@doi [\apj] {10.3847/1538-4357/aa6db4}, \href
  {http://adsabs.harvard.edu/abs/2017ApJ...841....5N} {841, 5}

\bibitem[\protect\citeauthoryear{{Planck Collaboration} et~al.,}{{Planck
  Collaboration} et~al.}{2016}]{PSZ2}
{Planck Collaboration} et~al., 2016, \mn@doi [\aap]
  {10.1051/0004-6361/201525823}, \href
  {https://ui.adsabs.harvard.edu/abs/2016A&A...594A..27P} {594, A27}

\bibitem[\protect\citeauthoryear{{Predehl} et~al.,}{{Predehl}
  et~al.}{2021}]{Predehl21}
{Predehl} P.,  et~al., 2021, \mn@doi [\aap] {10.1051/0004-6361/202039313},
  \href {https://ui.adsabs.harvard.edu/abs/2021A&A...647A...1P} {647, A1}

\bibitem[\protect\citeauthoryear{Raghunathan et~al.,}{Raghunathan
  et~al.}{2022}]{Raghunathan22}
Raghunathan S.,  et~al., 2022, \mn@doi [The Astrophysical Journal]
  {10.3847/1538-4357/ac4712}, 926, 172

\bibitem[\protect\citeauthoryear{{Reichardt} et~al.,}{{Reichardt}
  et~al.}{2013}]{Reichardt13}
{Reichardt} C.~L.,  et~al., 2013, \mn@doi [\apj] {10.1088/0004-637X/763/2/127},
  \href {https://ui.adsabs.harvard.edu/abs/2013ApJ...763..127R} {763, 127}

\bibitem[\protect\citeauthoryear{{Rykoff} et~al.,}{{Rykoff}
  et~al.}{2014}]{rykoff14}
{Rykoff} E.~S.,  et~al., 2014, \mn@doi [\apj] {10.1088/0004-637X/785/2/104},
  \href {https://ui.adsabs.harvard.edu/abs/2014ApJ...785..104R} {785, 104}

\bibitem[\protect\citeauthoryear{Schlafly, Meisner  \& Green}{Schlafly
  et~al.}{2019}]{unWISE3cat}
Schlafly E.~F.,  Meisner A.~M.,   Green G.~M.,  2019, \mn@doi [The
  Astrophysical Journal Supplement Series] {10.3847/1538-4365/aafbea}, 240, 30

\bibitem[\protect\citeauthoryear{{Sheldon}}{{Sheldon}}{2014}]{Sheldon14}
{Sheldon} E.~S.,  2014, \mn@doi [\mnras] {10.1093/mnrasl/slu104}, \href
  {https://ui.adsabs.harvard.edu/abs/2014MNRAS.444L..25S} {444, L25}

\bibitem[\protect\citeauthoryear{{Vanderlinde} et~al.,}{{Vanderlinde}
  et~al.}{2010}]{Vanderlinde10}
{Vanderlinde} K.,  et~al., 2010, \mn@doi [\apj] {10.1088/0004-637X/722/2/1180},
  \href {https://ui.adsabs.harvard.edu/abs/2010ApJ...722.1180V} {722, 1180}

\bibitem[\protect\citeauthoryear{{Wen} \& {Han}}{{Wen} \& {Han}}{2013}]{Wen13}
{Wen} Z.~L.,  {Han} J.~L.,  2013, \mn@doi [\mnras] {10.1093/mnras/stt1581},
  \href {http://adsabs.harvard.edu/abs/2013MNRAS.436..275W} {436, 275}

\bibitem[\protect\citeauthoryear{{Wright} et~al.,}{{Wright}
  et~al.}{2010}]{WISE}
{Wright} E.~L.,  et~al., 2010, \mn@doi [\aj] {10.1088/0004-6256/140/6/1868},
  \href {https://ui.adsabs.harvard.edu/abs/2010AJ....140.1868W} {140, 1868}

\bibitem[\protect\citeauthoryear{{de Haan} et~al.,}{{de Haan}
  et~al.}{2016}]{dehaan16}
{de Haan} T.,  et~al., 2016, \mn@doi [\apj] {10.3847/0004-637X/832/1/95}, \href
  {https://ui.adsabs.harvard.edu/abs/2016ApJ...832...95D} {832, 95}

\bibitem[\protect\citeauthoryear{{de Jong}, {Verdoes Kleijn}, {Kuijken}  \&
  {Valentijn}}{{de Jong} et~al.}{2013}]{deJong13}
{de Jong} J. T.~A.,  {Verdoes Kleijn} G.~A.,  {Kuijken} K.~H.,   {Valentijn}
  E.~A.,  2013, \mn@doi [Experimental Astronomy] {10.1007/s10686-012-9306-1},
  \href {https://ui.adsabs.harvard.edu/abs/2013ExA....35...25D} {35, 25}

\makeatother
\end{thebibliography}


\appendix

\section*{Data Availability}
The catalog will be made available as online supplement as well as on CDS.

\section*{Affiliations}
\noindent
{\it
$^{1}$ University Observatory, Faculty of Physics, Ludwig-Maximilians-Universit\"at, Scheinerstr. 1, 81679 Munich, Germany\\
$^{2}$ Max Planck Institute for Extraterrestrial Physics, Giessenbachstrasse, 85748 Garching, Germany\\
$^{3}$ Laborat\'orio Interinstitucional de e-Astronomia - LIneA, Rua Gal. Jos\'e Cristino 77, Rio de Janeiro, RJ - 20921-400, Brazil\\
$^{4}$ Kavli Institute for Particle Astrophysics and Cosmology, Stanford University, 452 Lomita Mall, Stanford, CA 94305, USA\\
$^{5}$ Department of Physics, Stanford University, 382 Via Pueblo Mall, Stanford, CA 94305, USA\\
$^{6}$ SLAC National Accelerator Laboratory, 2575 Sand Hill Road, Menlo Park, CA 94025, USA\\
$^{7}$ Department of Physics, University of Michigan, Ann Arbor, MI 48109, USA\\
$^{8}$ School of Physics, University of Melbourne, Parkville, VIC 3010, Australia\\
$^{9}$ Center for Astrophysics \textbar\ Harvard \& Smithsonian, Cambridge MA 02138, USA\\
$^{10}$ Institute of Cosmology and Gravitation, University of Portsmouth, Portsmouth, PO1 3FX, UK\\
$^{11}$ Kavli Institute for Astrophysics and Space Research, Massachusetts Institute of Technology, 77 Massachusetts Avenue, Cambridge, MA~02139, USA\\
$^{12}$ Department of Physics, University of Cincinnati, Cincinnati, OH 45221, USA\\
$^{13}$ Department of Astronomy and Astrophysics, University of Chicago, Chicago, IL 60637, USA\\
$^{14}$ Kavli Institute for Cosmological Physics, University of Chicago, Chicago, IL 60637, USA\\
$^{15}$ Fermi National Accelerator Laboratory, Batavia, IL 60510-0500, USA\\
$^{16}$ HEP Division, Argonne National Laboratory, Argonne, IL 60439, USA\\
$^{17}$ Department of Physics and Astronomy, University of Missouri, 5110 Rockhill Road, Kansas City, MO 64110, USA\\
$^{18}$ Department of Physics \& Astronomy, University College London, Gower Street, London, WC1E 6BT, UK\\
$^{19}$ Max Planck Institute for Extraterrestrial Physics, Giessenbachstr.\ 1, 85748 Garching, Germany\\
$^{20}$ Kavli Institute for Particle Astrophysics \& Cosmology, P. O. Box 2450, Stanford University, Stanford, CA 94305, USA\\
$^{21}$ SLAC National Accelerator Laboratory, Menlo Park, CA 94025, USA\\
$^{22}$ Institute of Cosmology \& Gravitation, University of Portsmouth, Dennis Sciama Building, Portsmouth, PO1 3FX, UK\\
$^{23}$ Department of Physics, University of Chicago, Chicago, IL 60637, USA\\
$^{24}$ Enrico Fermi Institute, University of Chicago, Chicago, IL 60637, USA\\
$^{25}$ Instituto de Astrofisica de Canarias, E-38205 La Laguna, Tenerife, Spain\\
$^{26}$ Universidad de La Laguna, Dpto. Astrofisica, E-38206 La Laguna, Tenerife, Spain\\
$^{27}$ Institut de F\'{\i}sica d'Altes Energies (IFAE), The Barcelona Institute of Science and Technology, Campus UAB, 08193 Bellaterra (Barcelona) Spain\\
$^{28}$ Jodrell Bank Center for Astrophysics, School of Physics and Astronomy, University of Manchester, Oxford Road, Manchester, M13 9PL, UK\\
$^{29}$ University of Nottingham, School of Physics and Astronomy, Nottingham NG7 2RD, UK\\
$^{30}$ Astronomy Unit, Department of Physics, University of Trieste, via Tiepolo 11, 34131 Trieste, Italy\\
$^{31}$ INAF - Osservatorio Astronomico di Trieste, via G. B. Tiepolo 11, 34143 Trieste, Italy\\
$^{32}$ IFPU - Institute for Fundamental Physics of the Universe, Via Beirut 2, 34014 Trieste, Italy\\
$^{33}$ California Institute of Technology, Pasadena, CA 91125, USA\\
$^{34}$ Hamburger Sternwarte, Universit\"{a}t Hamburg, Gojenbergsweg 112, 21029 Hamburg, Germany\\
$^{35}$ School of Mathematics and Physics, University of Queensland,  Brisbane, QLD 4072, Australia\\
$^{36}$ Centro de Investigaciones Energ\'eticas, Medioambientales y Tecnol\'ogicas (CIEMAT), Madrid, Spain\\
$^{37}$ Department of Physics, IIT Hyderabad, Kandi, Telangana 502285, India\\
$^{38}$ High Energy Accelerator Research Organization (KEK), Tsukuba, Ibaraki 305-0801, Japan\\
$^{39}$ Department of Physics, McGill University, Montreal, Quebec H3A 2T8, Canada\\
$^{40}$ Canadian Institute for Advanced Research, CIFAR Program in Cosmology and Gravity, Toronto, ON, M5G 1Z8, Canada\\
$^{41}$ Institute of Theoretical Astrophysics, University of Oslo. P.O. Box 1029 Blindern, NO-0315 Oslo, Norway\\
$^{42}$ Fermi National Accelerator Laboratory, P. O. Box 500, Batavia, IL 60510, USA\\
$^{43}$ European Southern Observatory, Karl-Schwarzschild-Strasse 2, 85748 Garching, Germany\\
$^{44}$ Department of Astronomy, University of Florida, Gainesville, FL 32611, USA\\
$^{45}$ Universit\"at Innsbruck, Institut f\"ur Astro- und Teilchenphysik, Technikerstr. 25/8, 6020 Innsbruck, Austria\\
$^{46}$ Center for Astrophysical Surveys, National Center for Supercomputing Applications, 1205 West Clark St., Urbana, IL 61801, USA\\
$^{47}$ Department of Astronomy, University of Illinois at Urbana-Champaign, 1002 W. Green Street, Urbana, IL 61801, USA\\
$^{48}$ Department of Astrophysical and Planetary Sciences and Department of Physics, University of Colorado, Boulder, CO 80309, USA\\
$^{49}$ Santa Cruz Institute for Particle Physics, Santa Cruz, CA 95064, USA\\
$^{50}$ Department of Physics, University of California, Berkeley, CA 94720, USA\\
$^{51}$ Center for Cosmology and Astro-Particle Physics, The Ohio State University, Columbus, OH 43210, USA\\
$^{52}$ Department of Physics, The Ohio State University, Columbus, OH 43210, USA\\
$^{53}$ University of Chicago, Chicago, IL 60637, USA\\
$^{54}$ Center for Astrophysics $\vert$ Harvard \& Smithsonian, 60 Garden Street, Cambridge, MA 02138, USA\\
$^{55}$ Department of Physics, University of California, Davis, CA 95616, USA\\
$^{56}$ High Energy Physics Division, Argonne National Laboratory, Lemont, IL 60439, USA\\
$^{57}$ Physics Division, Lawrence Berkeley National Laboratory, Berkeley, CA 94720, USA\\
$^{58}$ Centre for Extragalactic Astronomy, Durham University, South Road, Durham DH1 3LE, UK\\
$^{59}$ Institute for Computational Cosmology, Durham University, South Road, Durham DH1 3LE, UK\\
$^{60}$ Steward Observatory, University of Arizona, 933 North Cherry Avenue, Tucson, AZ 85721, USA\\
$^{61}$ George P. and Cynthia Woods Mitchell Institute for Fundamental Physics and Astronomy, and Department of Physics and Astronomy, Texas A\&M University, College Station, TX 77843,  USA\\
$^{62}$ Centro de Investigaciones Energ\'etches, Medioambientales y Tecnol\'ogicas (CIEMAT), Madrid, Spain\\
$^{63}$ Instituci\'o Catalana de Recerca i Estudis Avan\c{c}ats, E-08010 Barcelona, Spain\\
$^{64}$ Observat\'orio Nacional, Rua Gal. Jos\'e Cristino 77, Rio de Janeiro, RJ - 20921-400, Brazil\\
$^{65}$ Department of Physics, University of Minnesota, Minneapolis, MN 55455, USA\\
$^{66}$ Department of Physics and Astronomy, Pevensey Building, University of Sussex, Brighton, BN1 9QH, UK\\
$^{67}$ Center for Astrophysics \textbar\ Harvard \& Smithsonian, 60 Garden Street, Cambridge, MA 02138, USA\\
$^{68}$ Physics Department, Center for Education and Research in Cosmology and Astrophysics, Case Western Reserve University, Cleveland, OH 44106, USA\\
$^{69}$ Department of Physics, Case Western Reserve University, Cleveland, OH 44106, USA\\
$^{70}$ Department of Physics, Brookhaven National Laboratory, Upton, NY 11973, USA\\
$^{71}$ Universit\'e Paris-Saclay, CNRS, Institut d'Astrophysique Spatiale, 91405, Orsay, France\\
$^{72}$ INFN - National Institute for Nuclear Physics, Via Valerio 2, I-34127 Trieste, Italy\\
$^{73}$ ICSC - Italian Research Center on High Performance Computing, Big Data and Quantum Computing, Italy\\
$^{74}$ Liberal Arts Department, School of the Art Institute of Chicago, Chicago, IL 60603, USA\\
$^{75}$ Department of Astronomy, University of Michigan, 1085 S. University Ave, Ann Arbor, MI 48109, USA\\
$^{76}$ School of Physics and Astronomy, University of Southampton,  Southampton, SO17 1BJ, UK\\
$^{77}$ Kavli Institute for Astrophysics and Space Research, Massachusetts Institute of Technology, 77 Massachusetts Avenue, Cambridge, MA 02139, USA\\
$^{78}$ Kavli Institute for Particle Astrophysics \& Cosmology, P. O. Box 2450, Stanford University, Stanford, CA 94305, USA\\
$^{79}$ Institute of Geophysics and Planetary Physics, Lawrence Livermore National Laboratory, Livermore, CA 94551, USA\\
$^{80}$ INAF - Osservatorio Astronomico di Brera, Via Brera 28, I-20121, Milano, Italy \& Via Bianchi 46, I-23807, Merate, Italy\\
$^{81}$ Computer Science and Mathematics Division, Oak Ridge National Laboratory, Oak Ridge, TN 37831\\
$^{82}$  \\
$^{83}$ Dunlap Institute for Astronomy \& Astrophysics, University of Toronto, 50 St George St, Toronto, ON, M5S 3H4, Canada\\
$^{84}$ Department of Astronomy \& Astrophysics, University of Toronto, 50 St George St, Toronto, ON, M5S 3H4, Canada\\
$^{85}$ Astronomy Department, University of Illinois at Urbana-Champaign, 1002 W.\ Green Street, Urbana, IL 61801, USA\\
$^{86}$ Department of Physics, University of Illinois Urbana-Champaign, 1110 W.\ Green Street, Urbana, IL 61801, USA\\
$^{87}$ Department of Physics and Astronomy, Stony Brook University, Stony Brook, NY 11794, USA\\
$^{88}$ Lawrence Berkeley National Laboratory, 1 Cyclotron Road, Berkeley, CA 94720, USA\\
}

\label{lastpage}
\end{document}